\documentclass[12pt]{article}
\usepackage{geometry}                
\usepackage{graphicx}
\usepackage{amssymb}
\usepackage{mathtools}
\usepackage{epstopdf}
\usepackage{verbatim}
\usepackage{color}
\usepackage{slashed}
\usepackage{bbold}
\usepackage{fullpage}
\usepackage{float}
\usepackage[caption=true]{subfig}
\usepackage{afterpage}
\usepackage[title]{appendix}
\usepackage[colorlinks, citecolor=red, linkcolor=blue]{hyperref}
\usepackage[square, comma, numbers, sort&compress]{natbib}

\newcommand{\beq}{\begin{equation}}
\newcommand{\eeq}{\end{equation}}
\newcommand{\bmat}{\begin{pmatrix}}
\newcommand{\emat}{\end{pmatrix}}
\newcommand{\bal}{\begin{align}}
\newcommand{\eal}{\end{align}}

\newcommand{\Order}{\mathcal{O}}

\newcommand{\sq}{\tilde{q}}

\newcommand{\mStL}{m_{\sq_{L,3}}}
\newcommand{\mSqL}{m_{\sq_{L,i}}}
\newcommand{\mStR}{m_{\tilde{t}_{R,3}}}
\newcommand{\mSbR}{m_{\tilde{b}_{R,3}}}
\newcommand{\mSuR}{m_{\tilde{u}_{R,i}}}
\newcommand{\mSdR}{m_{\tilde{d}_{R,i}}}
\newcommand{\mSeR}{m_{\tilde{e}_{R,i}}}
\newcommand{\mSlL}{m_{\tilde{\ell}_{L,i}}}

\newcommand{\MS}{\overline{MS}}
\newcommand{\DR}{\overline{DR}}
\newcommand{\mS}{\tilde{m}}

\newcommand{\gev}{{\, {\rm GeV}}}

\newcommand{\gsim}{\lower.7ex\hbox{$\;\stackrel{\textstyle>}{\sim}\;$}}
\newcommand{\lsim}{\lower.7ex\hbox{$\;\stackrel{\textstyle<}{\sim}\;$}}

\begin{document}

\begin{titlepage}

\pagestyle{empty}

\baselineskip=21pt
\rightline{\footnotesize MCTP-17-06}
\vskip 0.6in

\begin{center}

{\large {\bf High-scale Supersymmetry, the Higgs Mass and Gauge Unification}}

\vskip 0.4in

 {\bf Sebastian~A.~R.~Ellis} and 
 {\bf James~D.~Wells}

\vskip 0.3in

{\small {\it
{Michigan Center for Theoretical Physics (MCTP), \\ Department of Physics, University of Michigan \\Ann Arbor, MI 48109 USA}\\
\vspace{0.25cm}
{Deutsches Elektronen-Synchrotron DESY}\\ \it{Notkestra\ss e 85, D-22607 Hamburg, Germany}}}

\vskip 0.5in

{\bf Abstract}

\end{center}

\baselineskip=18pt \noindent


{\small

Suppressing naturalness concerns, we discuss the compatibility requirements of high-scale supersymmetry breaking with the Higgs boson mass constraint and gauge coupling unification. We find that to accommodate superpartner masses significantly greater than the electroweak scale, one must introduce large non-degeneracy factors. These factors are enumerated, and implications for the allowed forms of supersymmetry breaking are discussed. We find that superpartner masses of arbitrarily high values are allowed for suitable values of $\tan\beta$ and the non-degeneracy factors. We also compute the large, but viable, threshold corrections that would be necessary at the unification scale for exact gauge coupling unification. Whether or not high-scale supersymmetry can be realised in this context is highly sensitive to the precise value of the top quark Yukawa coupling, highlighting the importance of future improvements in the top quark mass measurement.
}


\vskip 0.5in

{\small \leftline{May 31, 2017}}

\end{titlepage}

\newpage

\tableofcontents


\section{Introduction}

The discovery of the Higgs boson at the Large Hadron Collider (LHC) \cite{Aad:2012tfa, Chatrchyan:2012xdj} marked the culmination of the Standard Model (SM) as an effective theory of the electroweak scale. All properties measured so far are consistent with a simple Standard Model Higgs boson of mass 125 GeV. In the supersymmetric context this is consistent with the decoupling of the superpartner mass scale, where the lightest CP even Higgs eigenstate is SM-like and the others are heavy and inaccessible to the LHC. 
In view of the constraints on the supersymmetric spectra imposed by the measured Higgs mass, and the non-observation of superpartners at the weak scale, we discuss the possibility that supersymmetry exists at higher scales. This is uncomfortable from the perspective of naturalness, and we do not discuss it further here.

The next question to ask from a supersymmetry (SUSY) point of view is what does the measured Higgs boson mass imply for the superpartner spectrum. Previous studies have indicated that in order for SUSY to be reconciled with the observed Higgs mass, there are constraints from the requirement that unification at the high scale still occur \cite{Hall:2013eko, Hebecker:2014uaa, Hall:2014vga, Staub:2017jnp}. There are also experimental lower limits \cite{ATLAS-CONF-2017-022, ATLAS-CONF-2017-030, Sirunyan:2017cwe, Sirunyan:2017kqq}, and theoretical studies of upper limits on the scale of superpartners \cite{Carena:1995bx, Haber:1996fp, Bernal:2007uv, Giudice:2011cg, Ellis:2012aa, Cao:2012fz,Ibanez:2013gf,Djouadi:2013vqa, Bagnaschi:2014rsa, Hebecker:2014uaa, Lee:2015uza, Vega:2015fna}. However, for the most part, these studies have been concerned with determining approximate limits on the SUSY breaking scale in setups where the SUSY spectrum is close to degenerate. 

In this paper we show that the superpartner non-degeneracy is just as important in assessing the ability to accommodate the 125 GeV Higgs boson mass as is the overall scale of superpartner masses. We show below that, even when the scale of supersymmetry is orders of magnitude beyond the weak scale, any characteristic superpartner mass can fit the 125 GeV mass. Furthermore, we show that this can be consistent with exact gauge coupling unification. However, not any SUSY theory will do: requirements on non-degeneracy have strong implications for the type of supersymmetry breaking that is allowed, which we discuss at the end of the paper.

The paper is organised as follows. In Section \ref{Setup.SEC}, we discuss the matching condition for the SM to the SUSY theory as a function of the matching scale and $\tan\beta$ at one-loop order. We comment on the size of the necessary threshold corrections to the Higgs quartic coupling for matching to occur. In Section \ref{Corrections.SEC}, we analyse how the various one-loop corrections to the SUSY Higgs quartic coupling compare with one another. We then explain how we set up our analysis of non-degenerate spectra, and present results for various different choices of characteristic SUSY scales and values of $\tan\beta$. In Section \ref{Examples.SEC}, we discuss specific example spectra at each choice of SUSY scale, where matching to the SM has been achieved. Finally in Section \ref{GCU.SEC}, we discuss the implications for gauge coupling unification. We summarise our findings in Section \ref{Conclusion.SEC}.

\section{Higgs self-coupling matching}
\label{Setup.SEC}

The Standard Model Higgs potential is
\beq
V(H)=\frac{\lambda_H}{2} \left( |H|^2-v^2 \right)^2
\eeq
which implies that after symmetry breaking, the physical propagating Higgs boson field $h$ has mass $m_h^2=2\lambda_H v^2$ at tree level, where $v\simeq 174\gev$. 

In this paper we assume that the SM is a low-energy effective theory of a minimal supersymmetric (MSSM) model that was integrated out at a scale $\mS$, which is characteristic of the superpartner masses. 
The well-known tree-level matching condition for the scalar quartic coupling at the SUSY scale is
\begin{align}
\lambda^{tree}_H(\mS) = \frac{1}{4}\left( g_2^2(\mS) + \frac{3}{5}g_1^2(\mS) \right) \cos^2 2\beta \ ,
\label{TreeLambda.EQ}
\end{align}
where $\mS$ is the SUSY scale, $g_i$ are the $SU(2)$ and $U(1)_Y$ gauge couplings with the appropriate GUT normalisation and $\beta$ is conventionally the angle associated with the ratio of the vacuum expectation values of the two Higgs doublets $H_u$ and $H_d$. In our analysis, as in \cite{Bagnaschi:2014rsa}, the treatment of $\tan\beta$ requires extra care, and will be discussed below in Appendix \ref{TB.APP}. All the couplings above are in the $\MS$ scheme, which explicitly breaks supersymmetry. Therefore in the supersymmetric regime one must switch to the $\DR$ scheme, which will lead to finite corrections to the above relation.

This tree-level relation receives threshold corrections at the scale $\mS$, which can be large if the superpartners are not precisely degenerate. Thus at one loop the matching condition for $\lambda$ becomes
\begin{align}
\lambda_H(\mS) = \lambda_H^{tree} (\mS) + \Delta\lambda_H^{(1)}  (\mS) \ ,
\label{1LLambda.EQ}
\end{align}
where $ \Delta\lambda_H^{(1)}  (\mS)$ incorporates the finite correction due to switching between the $\MS$ scheme in the SM phase and $\DR$ in the SUSY phase, as well as the corrections due to scalar and gaugino/higgsino loops. Expressions for these one-loop corrections are provided in Appendix~\ref{Scheme.APP}.

\begin{figure}[t]
\centering
\includegraphics[scale=0.6]{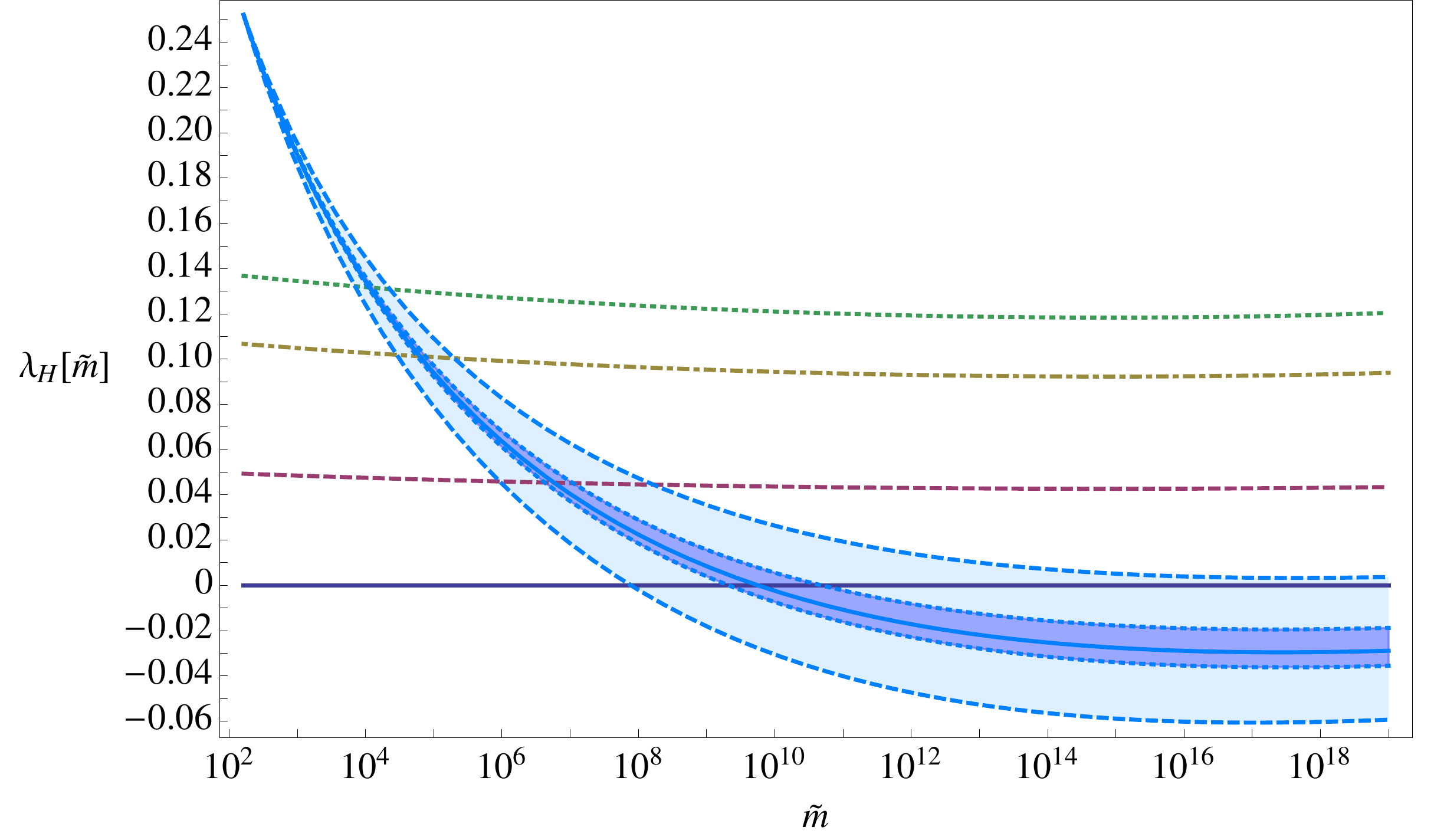}
\caption{Plot showing the running of $\lambda_H$ in the SM (solid blue line), with 3$\sigma$ contours corresponding to both the error in $\alpha_s(m_t)$ (inner dotted blue lines) and $y_t(m_t)$ (outer dashed blue lines). Also shown is the tree-level SUSY matching condition for values of $\tan \beta = 1, 2, 4, 50$ (dark blue solid, dashed purple, dot-dashed olive, dotted green).}
\label{LambdaRunning.FIG}
\end{figure}

Let us now analyse the conditions  required for matching of the SM $\lambda_H$ to the MSSM. In Fig. \ref{LambdaRunning.FIG}  we see how the SM and the tree-level MSSM values for $\lambda_H$ vary as a function of the scale $\mS$. Since our definition of $\tan\beta$ is valid at the input scale, which can be chosen to be any value of $\mS$, we do not include running of $\tan\beta$ here. Therefore the slight variation with $\mS$ of the MSSM tree-level relation for $\lambda_H$ is entirely due to the running of $g_{1,2}$, which we calculate to two-loop accuracy. For the RGE evolution of the SM $\lambda_H$, we use the partial three-loop results provided in \cite{Bagnaschi:2014rsa}, for greater accuracy. Our SM input values, with their corresponding errors, are taken from ref.~\cite{Agashe:2014kda}.

Of note is that from Fig. \ref{LambdaRunning.FIG}, the na\"ive maximum SUSY scale appears to be $\mS \simeq 10^{10}$ GeV, since that is where the SM central value of $\lambda_H$ crosses the absolute minimum tree-level MSSM value of $\lambda^{tree}_H = 0$. However, it is worth remarking that if one takes the value of the top quark Yukawa coupling as $y_t(m_t)_{central}-3\sigma_{y_t(m_t)}$, one finds that the SM value of $\lambda_H$ is always greater than 0. This highlights the importance of an accurate measurement of $m_t(m_t)$, both for a better understanding of how the Higgs mass can be matched onto a SUSY theory, and of course for its implications for electroweak vacuum stability.

\begin{figure}[t]
\centering
\includegraphics[scale=0.6]{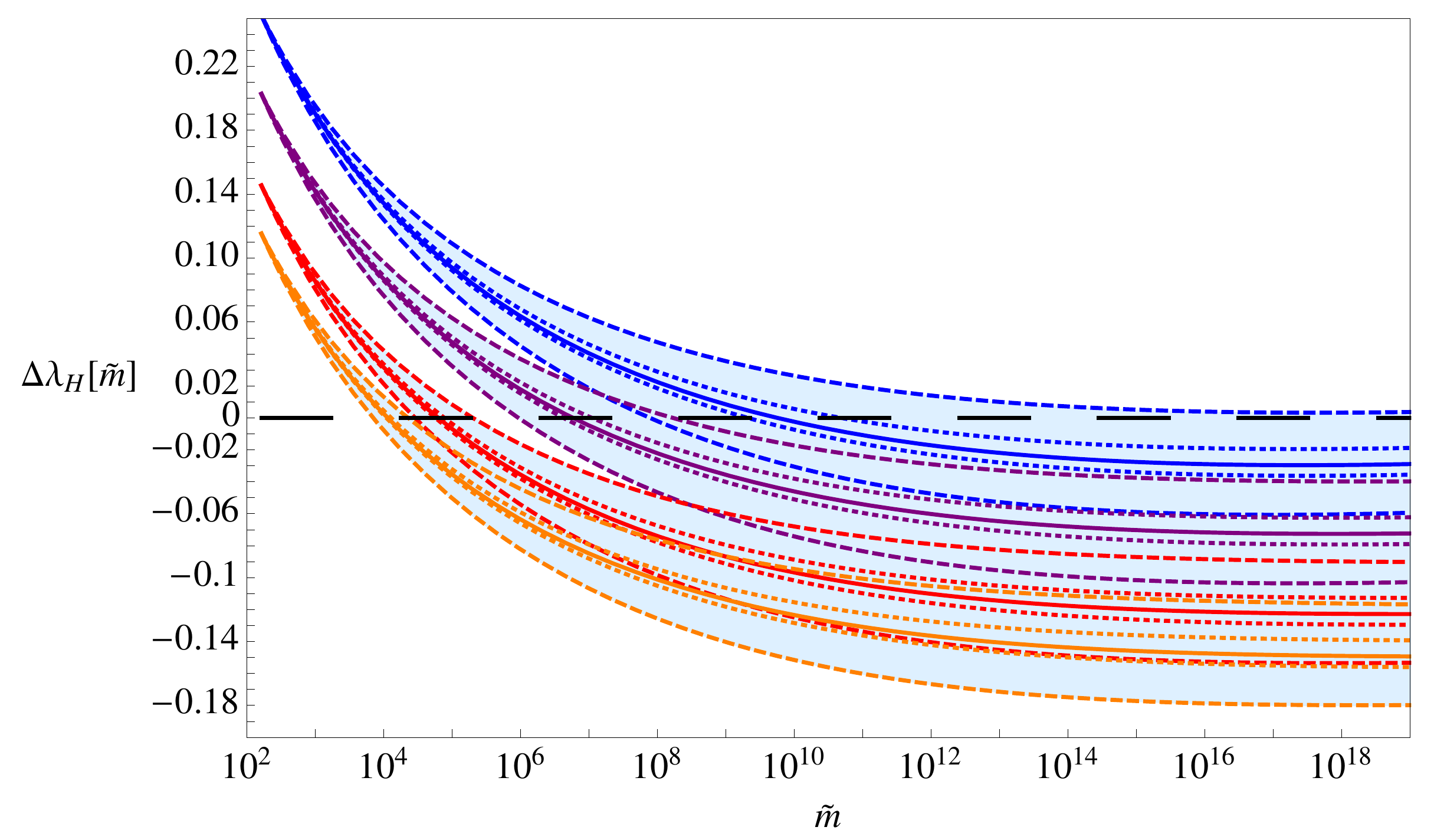}
\caption{Plot showing the required threshold corrections for the tree-level SUSY $\lambda^{tree}_H(\mS)$ to match the SM $\lambda^{SM}_H(\mS)$ at a given scale $\mS$. Again, 3$\sigma$ contours are shown, with the same definition as in Fig. \ref{LambdaRunning.FIG}. Shown is $\Delta \lambda^{req.}_H(\mS) = \lambda^{SM}_H(\mS) - \lambda^{tree}_H(\mS)$ for $\tan \beta = 1, 2, 4, 50$ (blue, purple, red, orange).}
\label{DeltaLambdaRunning.FIG}
\end{figure}

As a function of the scale $\mS$, we can define the required higher-order corrections to the SUSY tree-level value of $\lambda^{tree}_H$ needed to match the SM value at the scale $\mS$:
\begin{align}
\Delta\lambda^{req.}_H (\mS) = \lambda^{SM}_H (\mS) - \lambda^{tree}_H (\mS) \ .
\end{align}
These needed corrections are plotted in Fig.~\ref{DeltaLambdaRunning.FIG}.
Thus, at every scale $\mS$, there is a certain $\Delta\lambda^{req.}_H(\mS)$ that is required, which can then be compared with the one-loop threshold corrections $\Delta\lambda^{(1)}_H(\mS)$, for various SUSY spectra. We see from Fig.~\ref{DeltaLambdaRunning.FIG} that the required higher-order threshold corrections depend most significantly on the parameters $\tan\beta$ and $\mS$.

\begin{figure}[t]
\centering
\includegraphics[scale=0.6]{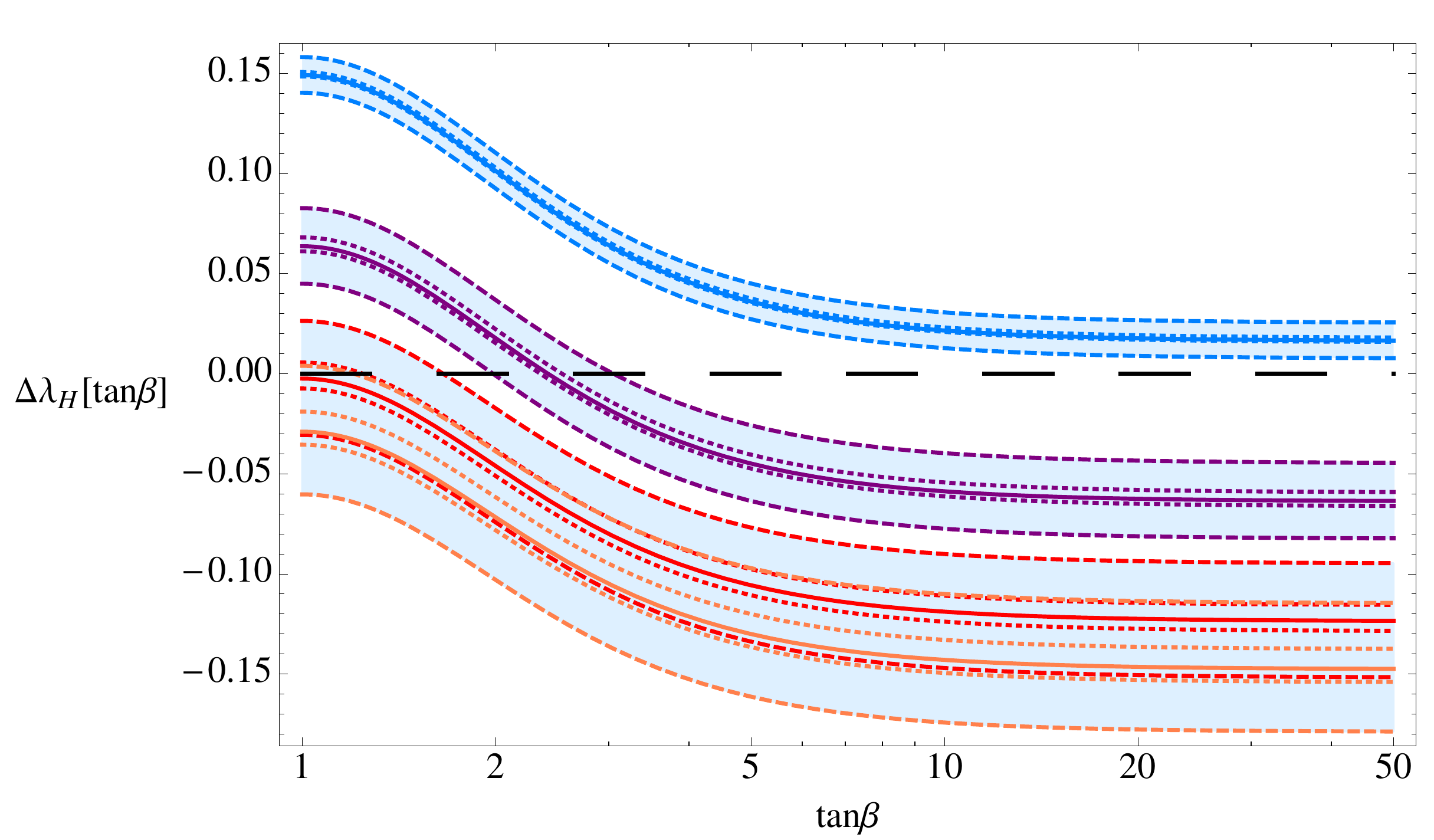}
\caption{Plot showing the required threshold corrections for the tree-level SUSY $\lambda^{tree}_H(\mS)$ to match the SM $\lambda^{SM}_H(\mS)$ as a function of $\tan \beta$. Again, 3$\sigma$ contours are shown, with the same definition as in Fig. (\ref{LambdaRunning.FIG}). Shown is $\Delta \lambda^{req.}_H(\mS) = \lambda^{SM}_H(\mS) - \lambda^{tree}_H(\mS)$ for $\mS = 5\times10^3,~10^6,~ 10^{10},~10^{16}$ GeV (blue, purple, red, orange).}
\label{DeltaLambdaPlotTB.FIG}
\end{figure}

 We now show, in Fig. \ref{DeltaLambdaPlotTB.FIG}, the required threshold corrections as a function of $\tan\beta$ for various different choices of $\mS$. From this figure we can see that for any given SUSY scale there is a naively preferred value of $\tan\beta$, corresponding to where the required threshold corrections are zero. For GUT-scale SUSY, since the required threshold corrections are always negative, there is no tree-level preferred value of $\tan\beta$. As one gets to large values of $\tan\beta\gtrsim 5$, we see that the required threshold corrections asymptote to a fixed value. This is due to the asymptotic behavior of $\cos^2\beta$ which appears in the SUSY tree-level matching condition, Eq.~\ref{TreeLambda.EQ}.

\section{Achieving sufficiently large threshold corrections}
\label{Corrections.SEC}

\begin{figure}[t!]
\centering
\includegraphics[scale=0.9]{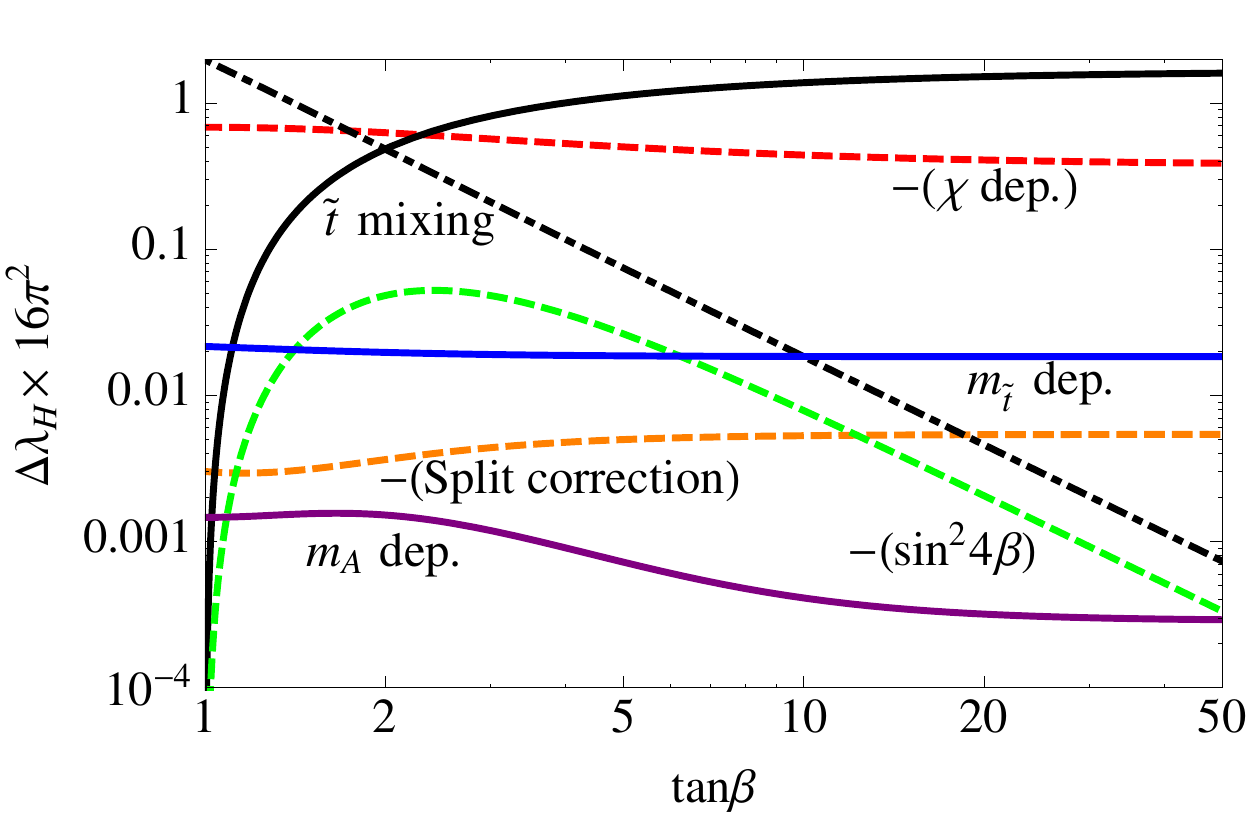}
\caption{Plot showing the variation of different parts of the threshold corrections $\Delta\lambda_H$ at one-loop, in units of $16\pi^2$, as a function of $\tan \beta$ with $\mS = 10^4 \ \gev$. Shown are the threshold corrections defined in Eqs. (\ref{ThirdGenPart.EQ})-(\ref{Gauginos1L.EQ}), divided into sub-components, with all superpartners chosen to be almost degenerate ($m_i = 1.01 \mS$), so that the ``$\log$"s are not all zero. In blue is the pure $\mStL,\mStR$ part, corresponding to the first three lines of Eq. (\ref{ThirdGenPart.EQ}). In black is the stop mixing part, corresponding to the last three lines of Eq. (\ref{ThirdGenPart.EQ}) (black dot-dashed corresponds to stop mixing with $A_t = 0$.) The 1st and 2nd generation squark, and all slepton generation contribution, corresponding to Eq. (\ref{OtherGenPart.EQ}) is smaller, and therefore is not shown. The part of the scalar corrections which is independent of the scalar masses, corresponding to Eq. (\ref{SinBeta.EQ}), is shown in green. The $m_A$-dependent part from Eq. (\ref{mAPart.EQ}) is shown in purple. The red line corresponds to the first four lines of Eq. (\ref{Gauginos1L.EQ}). The orange line shows a possible correction to the running of the gauge couplings in a Split SUSY setup, due to the Higgsino mass parameter $\mu$ being considerably lighter than the rest of the spectrum. All solid lines indicate positive corrections, while dashed lines indicate negative corrections.}
\label{DeltaLambdaTB.FIG}
\end{figure}

Now that we have determined that either positive or negative threshold corrections may be required to match the SM to the tree-level SUSY Higgs quartic coupling, we would like to investigate whether and how such corrections may be achieved in the MSSM. In Fig. \ref{DeltaLambdaTB.FIG}, we see how different contributions to the one-loop threshold corrections in the MSSM vary as a function of $\tan\beta$. We show this for an almost degenerate spectrum, so that we know the sign of the coefficient in front of the $\log$'s in Eqs. (\ref{ThirdGenPart.EQ}-\ref{Gauginos1L.EQ}). This allows us to see that in order to get overall negative threshold corrections, we need there to be a significant contribution from the gauginos/higgsinos, while minimising the contribution from the scalars. To achieve large positive threshold corrections (such as for low-scale SUSY with large $\tan\beta$), one should maximise stop mixing as expected.

Comparing how the different contributions to the threshold corrections vary as a function of the SUSY scale $\mS$, we find that all the contributions become smaller for larger $\mS$. Most contributions only vary a small amount, with for example the gaugino contributions given in Eq. (\ref{Gauginos1L.EQ}) decreasing such that

\beq
\Delta\lambda_H^{(1),-ino}\biggr\rvert_{10^{16} \gev}\sim 0.75 ~ \Delta\lambda_H^{(1),-ino}\biggr\rvert_{10^{3} \gev}\ ,
\eeq
with only small variation in the numerical factor as a function of $\tan\beta$.

The decrease at higher $\mS$ for both the stop contributions in the first three lines of Eq. (\ref{ThirdGenPart.EQ}) as well as the stop mixing contributions from the last three lines of Eq. (\ref{ThirdGenPart.EQ}) is more substantial, with
\begin{align}
&\Delta\lambda_H^{(1),\tilde{t}}\biggr\rvert_{10^{16} \gev} \sim [0.06,0.02] ~ \Delta\lambda_H^{(1),\tilde{t}}\biggr\rvert_{10^{3} \gev} \ , ~\tan\beta \in [1,50] \ ,
\end{align}

\begin{align}
&\Delta\lambda_H^{(1),\tilde{t}-mixing}\biggr\rvert_{10^{16} \gev} \sim [0.06,0.04] ~ \Delta\lambda_H^{(1),\tilde{t}-mixing}\biggr\rvert_{10^{3} \gev} \ , ~\tan\beta \in [1.01,50] \ ,
\end{align}
regardless of the choice of $A_t$, except for the special case where $|A_t -\mu \cot\beta|= \mS = \mStL = \mStR$, which results in $\Delta\lambda_H^{(1),~\tilde{t}-mixing} = 0$ for all values of $\mS$.

Having determined how large the threshold corrections must be in order to match the tree-level SUSY relation to the SM $\lambda_H$, as well as how various superpartners would contribute to the MSSM threshold corrections, we are now in a position to perform scans of SUSY spectra  to find solutions to the one-loop matching condition.

We study the one-loop matching for three different high SUSY scales, $\mS = 10^{6}$, $10^{10}$, and $10^{16}$ GeV. Each choice is motivated for different reasons. We also include the scale $\mS=5\times10^3$, since it is of interest for naturalness and due to the viability of the LSP as a dark matter candidate. The scale $\mS=10^6$ GeV is of interest as the scale of Split Supersymmetry \cite{Wells:2003tf, ArkaniHamed:2004fb, Giudice:2004tc, ArkaniHamed:2004yi, Wells:2004di}.
However, it is worth remarking that in typical Split Supersymmetry setups, $\mu$ and the gaugino masses are signifcantly lighter than $\mS$, so one must account for that separation with modified running of $\lambda_H$ between the various scales. In our analysis, we keep $\mu$ and the gaugino masses fairly close to the typical superpartner scale $\mS$. 
The choice of the intermediate scale of $\mS=10^{10}$ GeV is motivated because it corresponds to the scale which the tree-level matching condition suggests to be the na\"ive maximum SUSY scale. Finally, the choice of $\mS=10^{16}$ GeV is motivated by the possibility of associating SUSY breaking to GUT breaking. Having SUSY at the GUT scale is also interestingly compatible with recent proposals for ultra-heavy Gravitino Dark Matter \cite{Benakli:2017whb, Dudas:2017rpa}.

\begin{figure}[H]
\centering
\subfloat[$\tan\beta = 1$]{\includegraphics[width=0.5\textwidth]{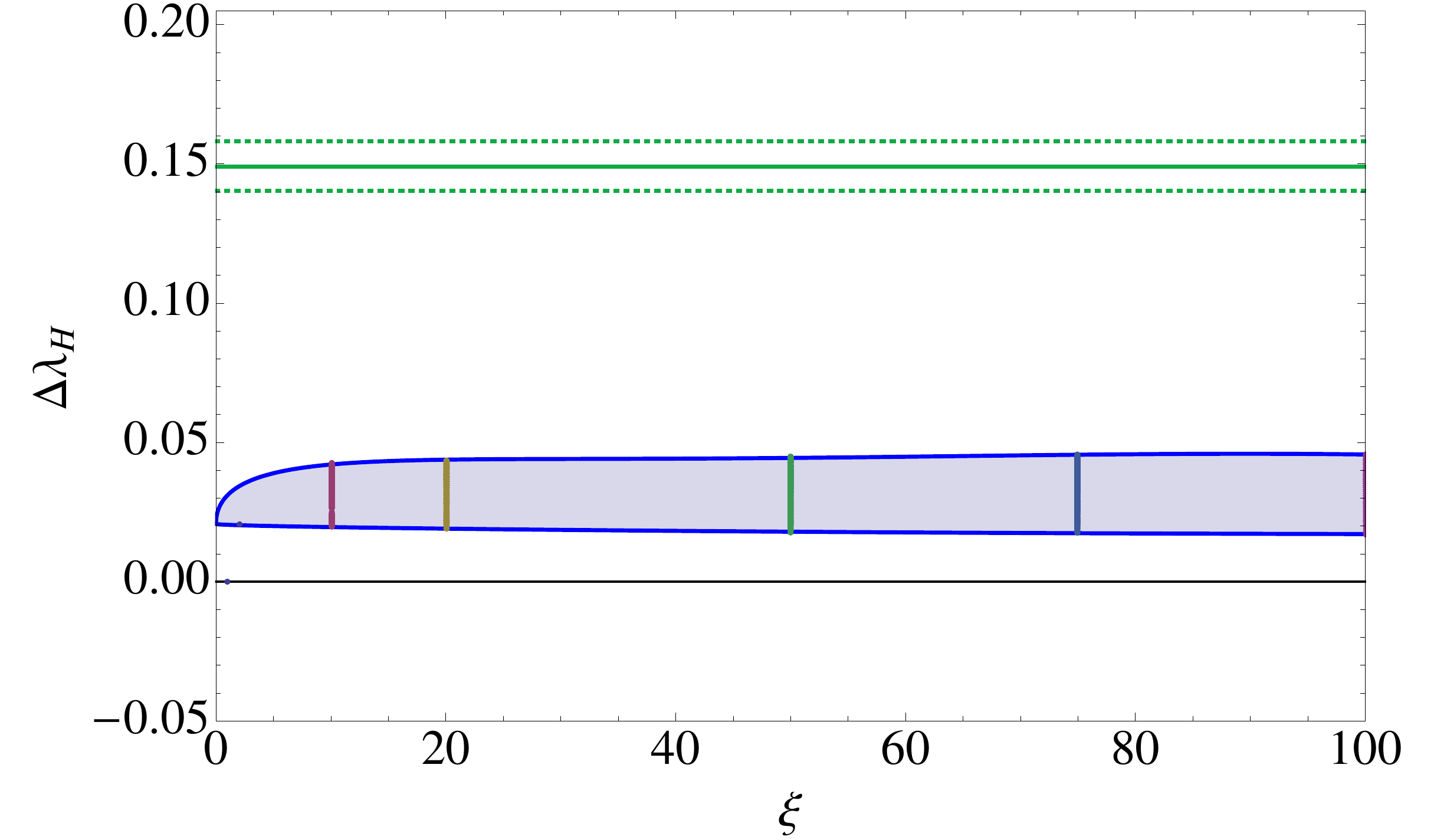}
}
\subfloat[$\tan\beta = 2$]{\includegraphics[width=0.5\textwidth]{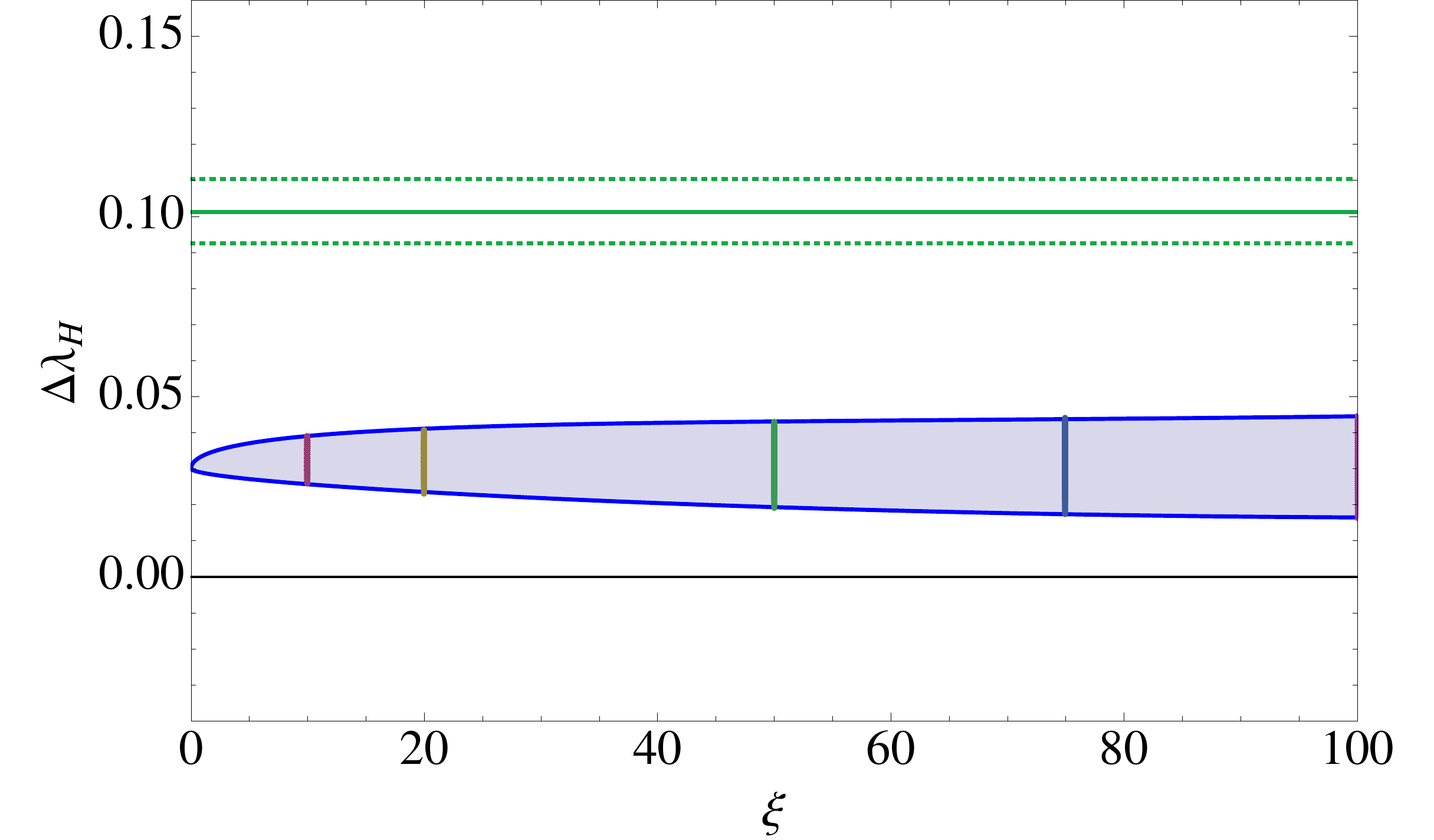}
}
\\
\subfloat[$\tan\beta = 4$]{\includegraphics[width=0.5\textwidth]{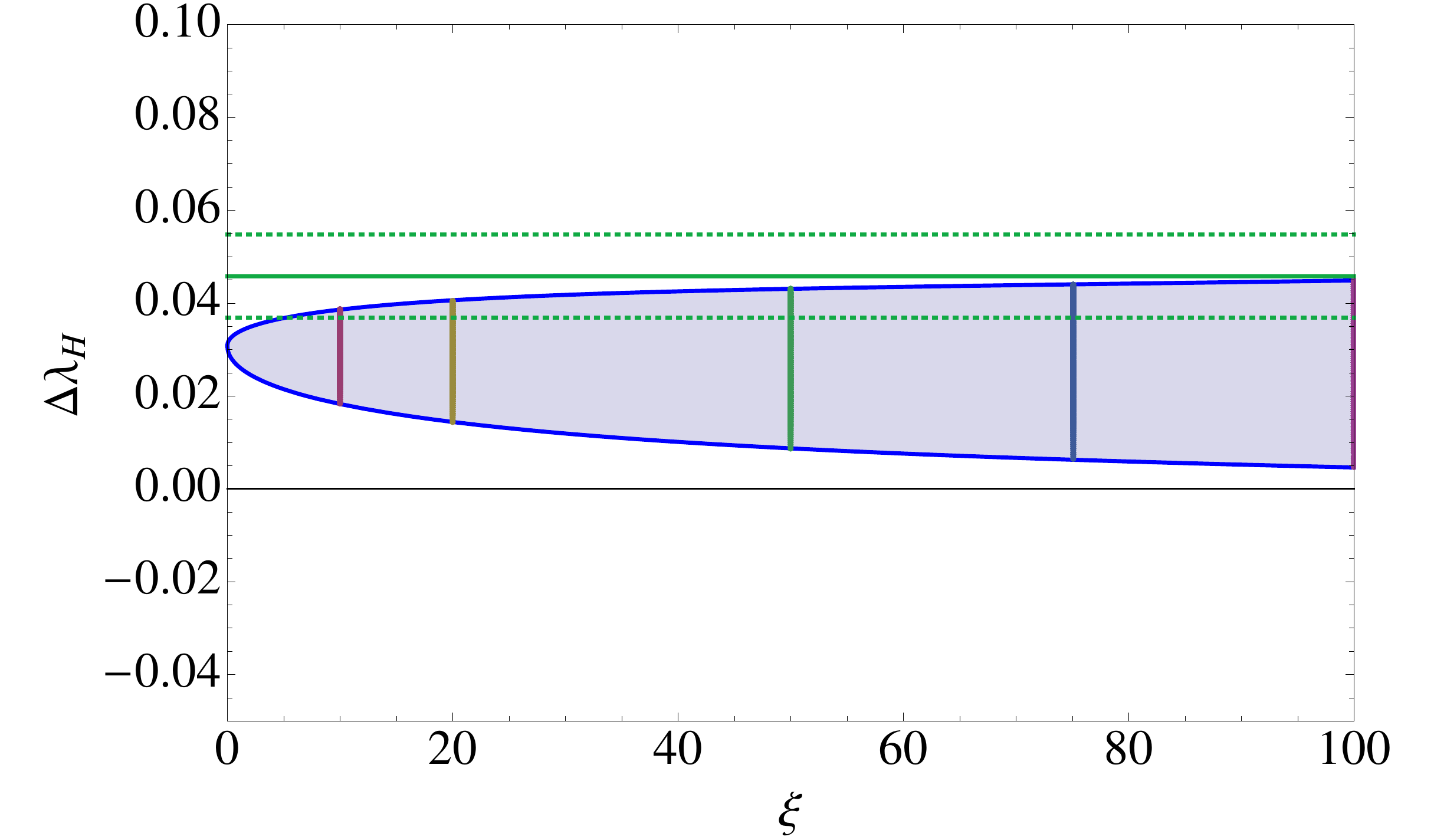}
}
\subfloat[$\tan\beta = 50$]{\includegraphics[width=0.5\textwidth]{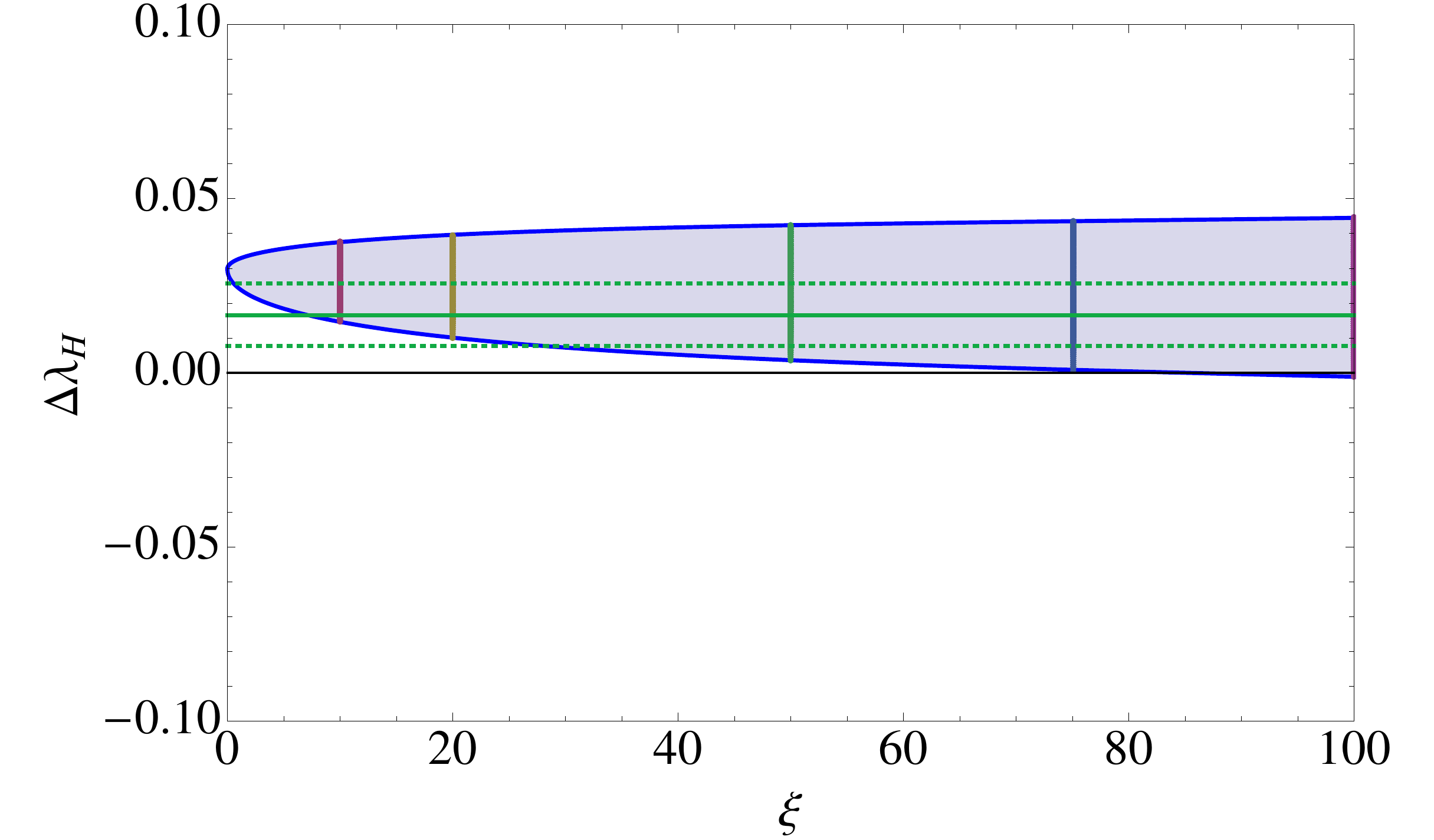}
}
\caption{
These plots show how the required corrections to $\lambda_H$ (solid green) compare to the upper and lower limits of the threshold corrections $\Delta\lambda$ obtained from a scan, at a SUSY scale $\mS = \mStL=\mStR = 5\times10^3$ GeV. 
The dashed green lines correspond to the $3\sigma$ upper and lower limits from $m_t$ uncertainty. 
Scanned independently are $ \mu,~ \{ \mSqL, \mSuR, \mSdR, \mSbR, \mSlL, \mSeR, m_A \},~ \{ M_a\}$ and $\tan\beta$. The top trilinear $A_t$ has been set to $\sqrt{6}\ \mS$. Each mass is allowed to vary up to $\xi=100$ relative to $\mS$, with scanned points shown here for $\xi=0,10,20,50,75$ and $100$. The four plots show the results for four separate choices of $\tan \beta$.
}
\label{TeVVary.FIG}
\end{figure}

\begin{figure}[H]
\centering
\subfloat[$\tan\beta = 1$]{\includegraphics[width=0.5\textwidth]{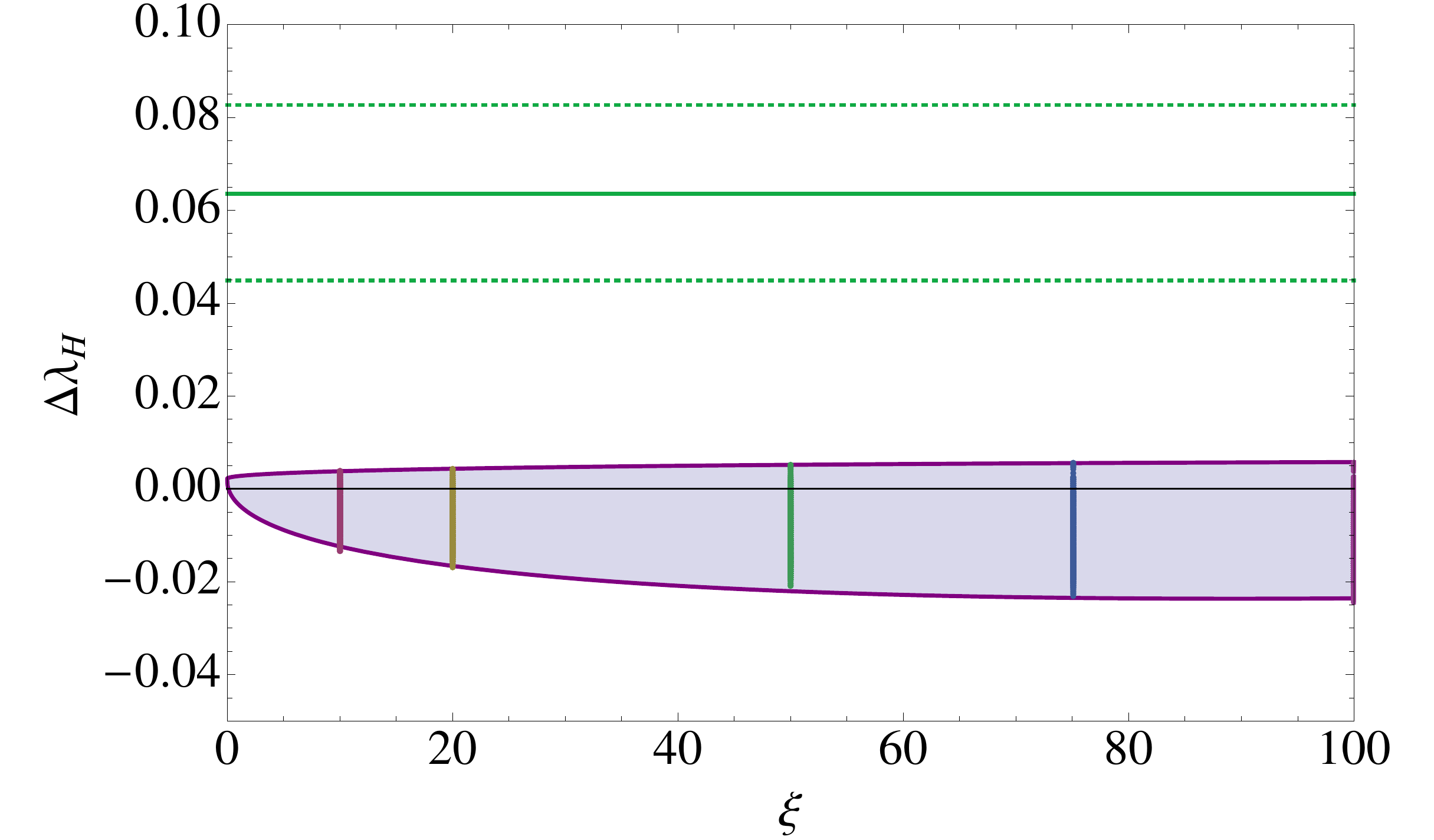}
}
\subfloat[$\tan\beta = 2$]{\includegraphics[width=0.5\textwidth]{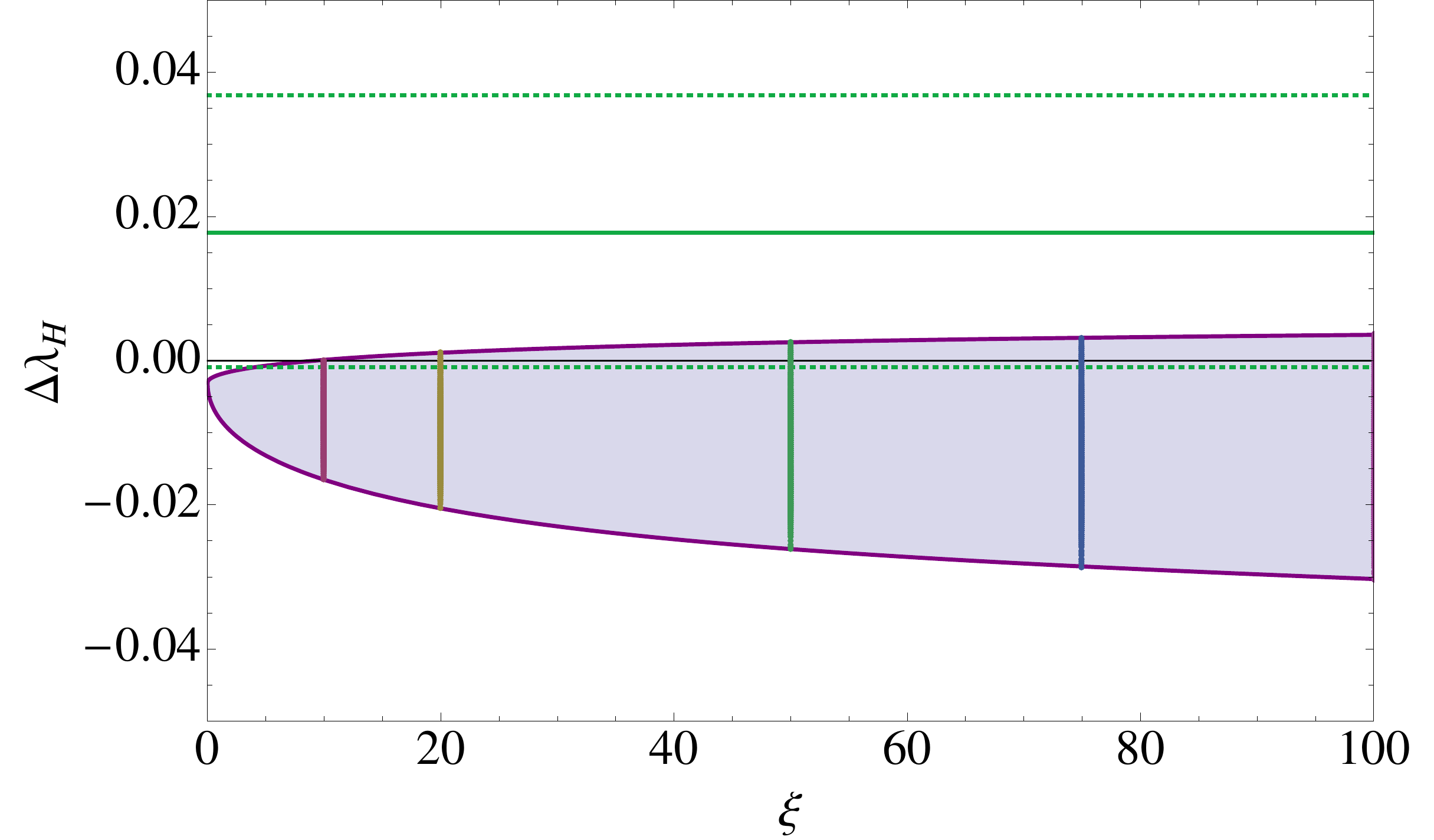}
}
\\
\subfloat[$\tan\beta = 4$]{\includegraphics[width=0.5\textwidth]{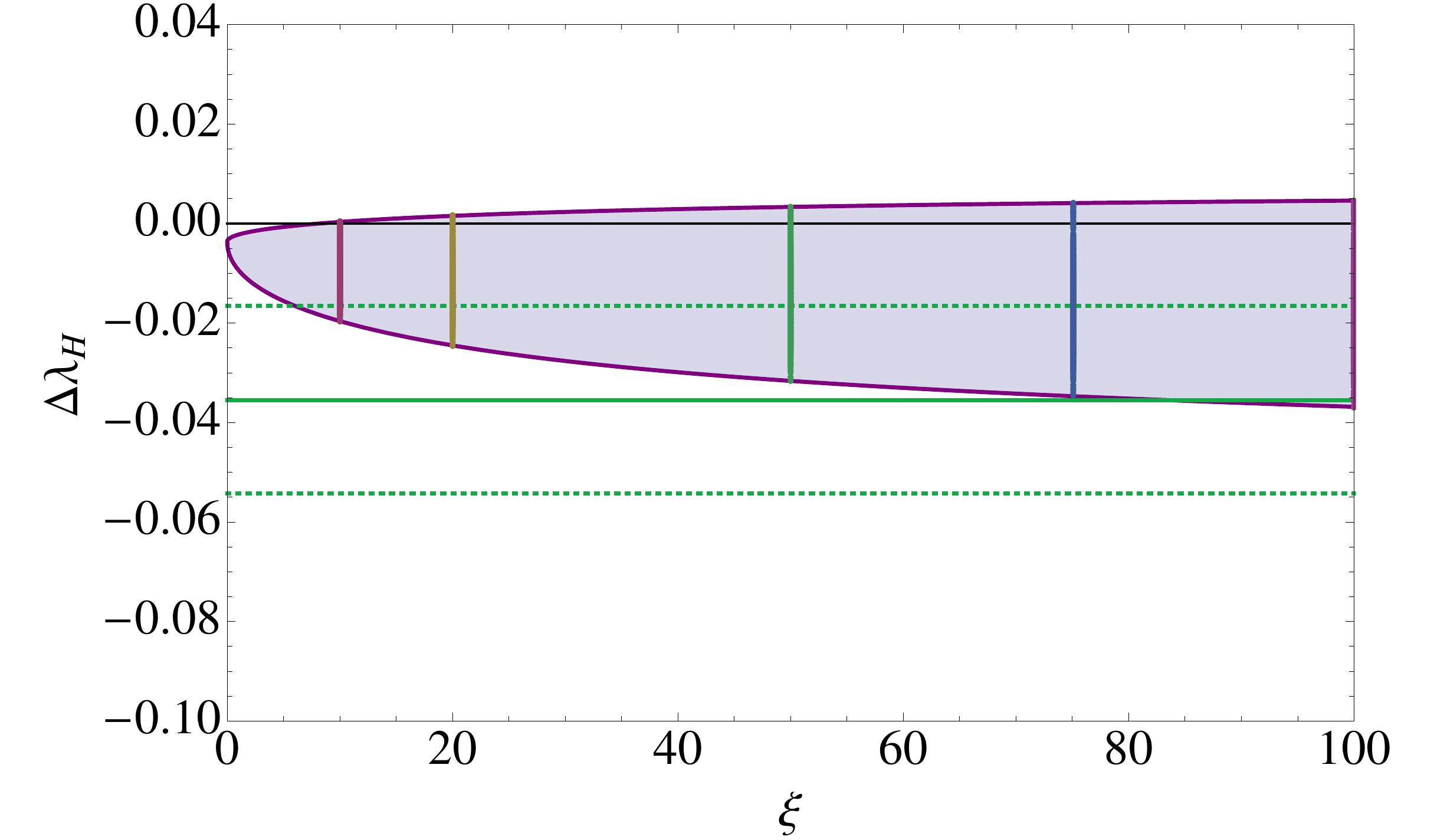}
}
\subfloat[$\tan\beta = 50$]{\includegraphics[width=0.5\textwidth]{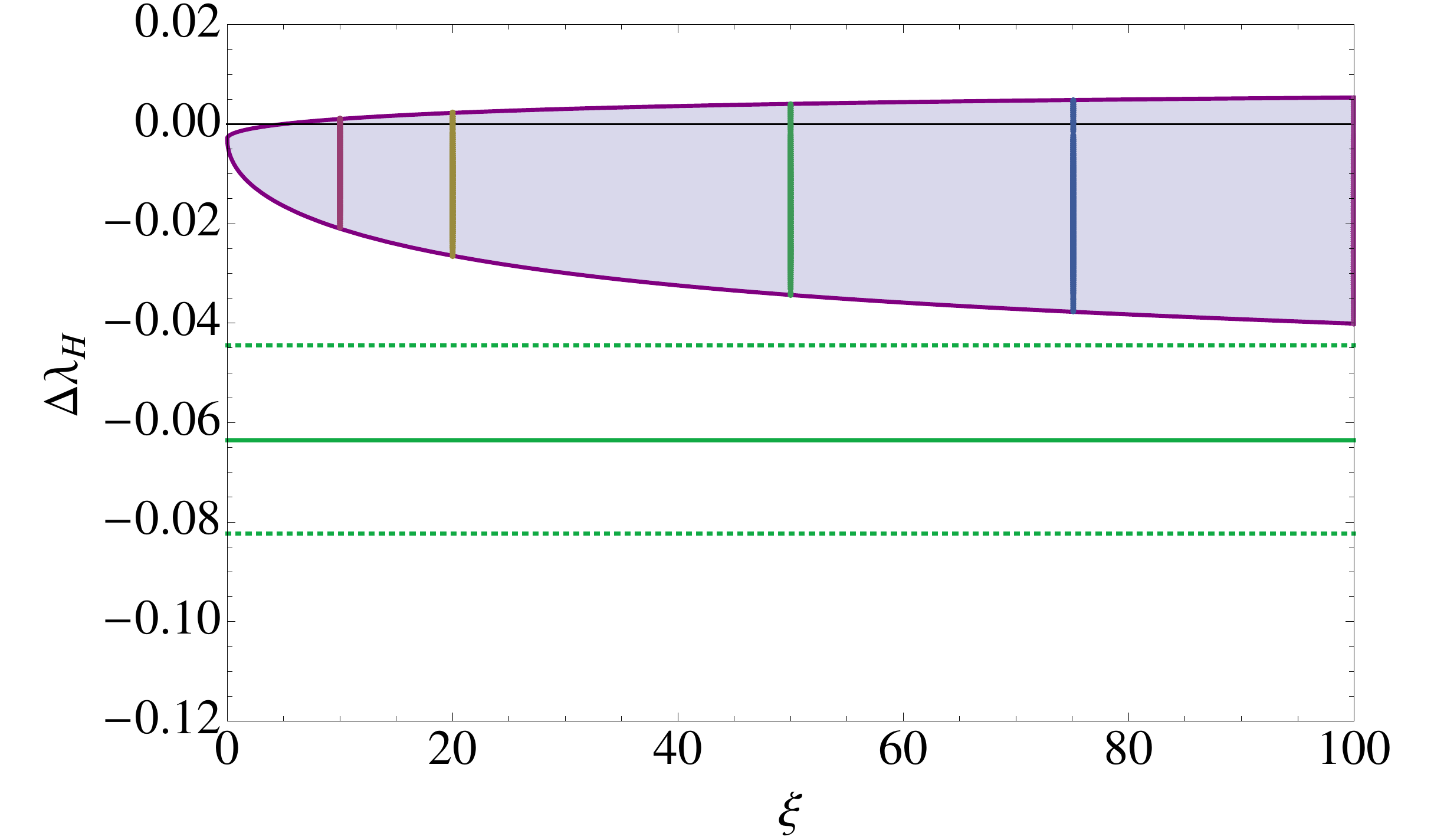}
}
\caption{
These plots show how the required corrections to $\lambda_H$ (solid green) compare to the upper and lower limits of the threshold corrections $\Delta\lambda$ obtained from a scan, at a SUSY scale $\mS = \mStL=\mStR = 10^6$ GeV. 
The dashed green lines correspond to the $3\sigma$ upper and lower limits from $m_t$ uncertainty.
Scanned independently are $ \mu,~ \{ \mSqL, \mSuR, \mSdR, \mSbR, \mSlL, \mSeR, m_A \},~ \{ M_a\}$ and $\tan\beta$. The top trilinear $A_t$ has been set to zero. Each mass is allowed to vary up to $\xi=100$ relative to $\mS$, with scanned points shown here for $\xi=0,10,20,50,75$ and $100$. The four plots show the results for four separate choices of $\tan \beta$.
}
\label{PeVVary.FIG}
\end{figure}

\begin{figure}[H]
\centering
\subfloat[$\tan\beta = 1$]{\includegraphics[width=0.5\textwidth]{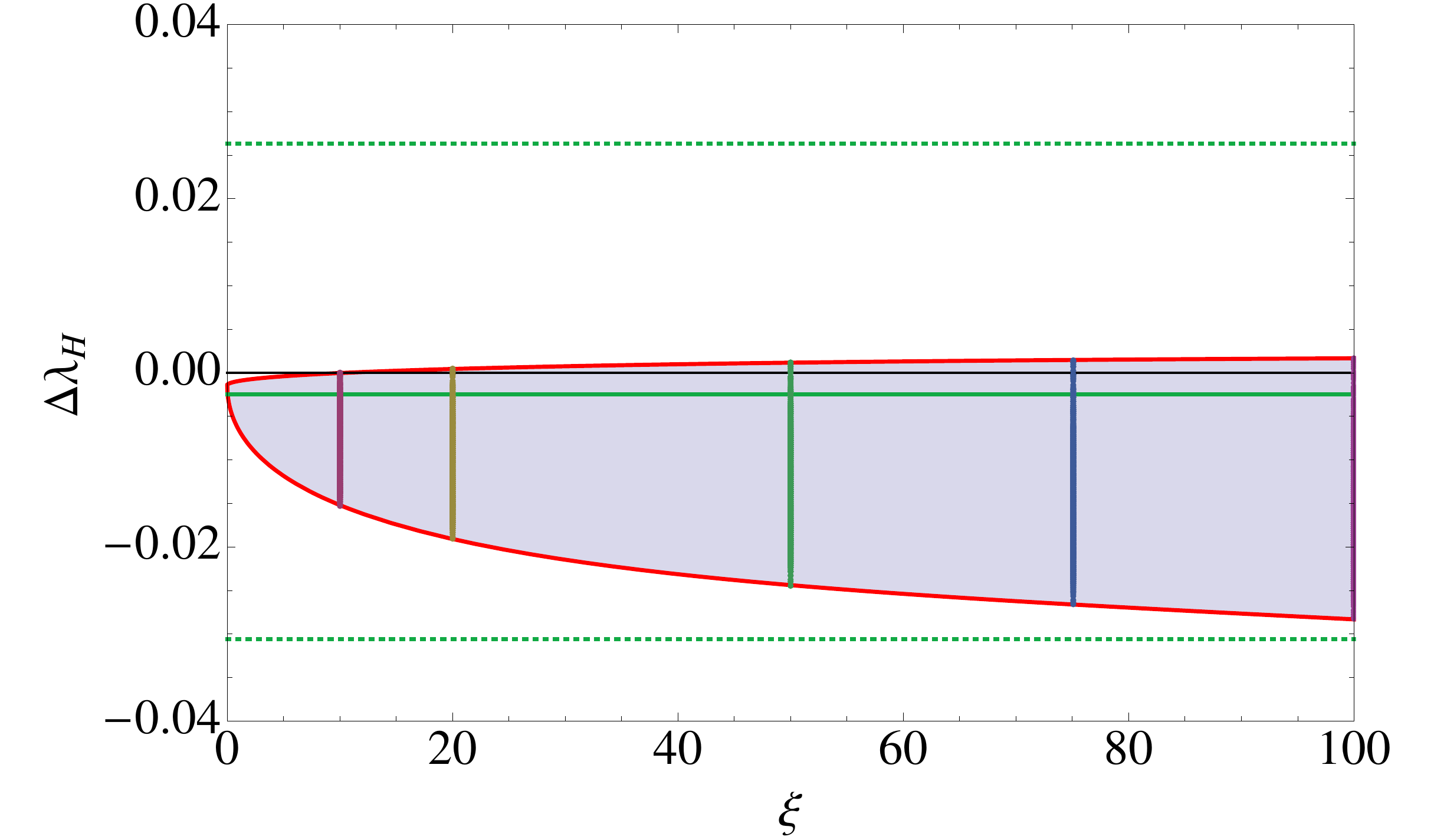}
}
\subfloat[$\tan\beta = 2$]{\includegraphics[width=0.5\textwidth]{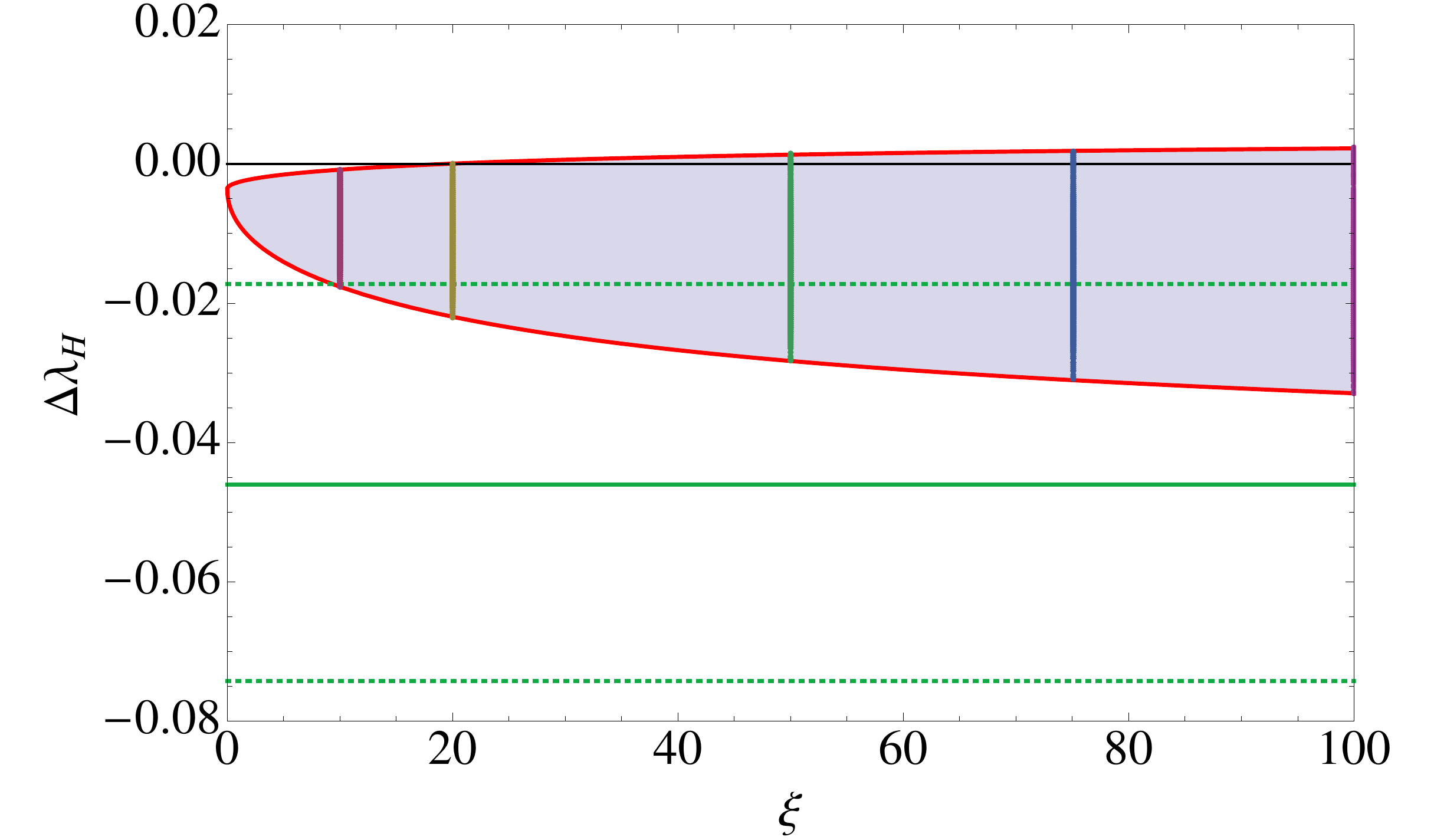}
}
\\
\subfloat[$\tan\beta = 4$]{\includegraphics[width=0.5\textwidth]{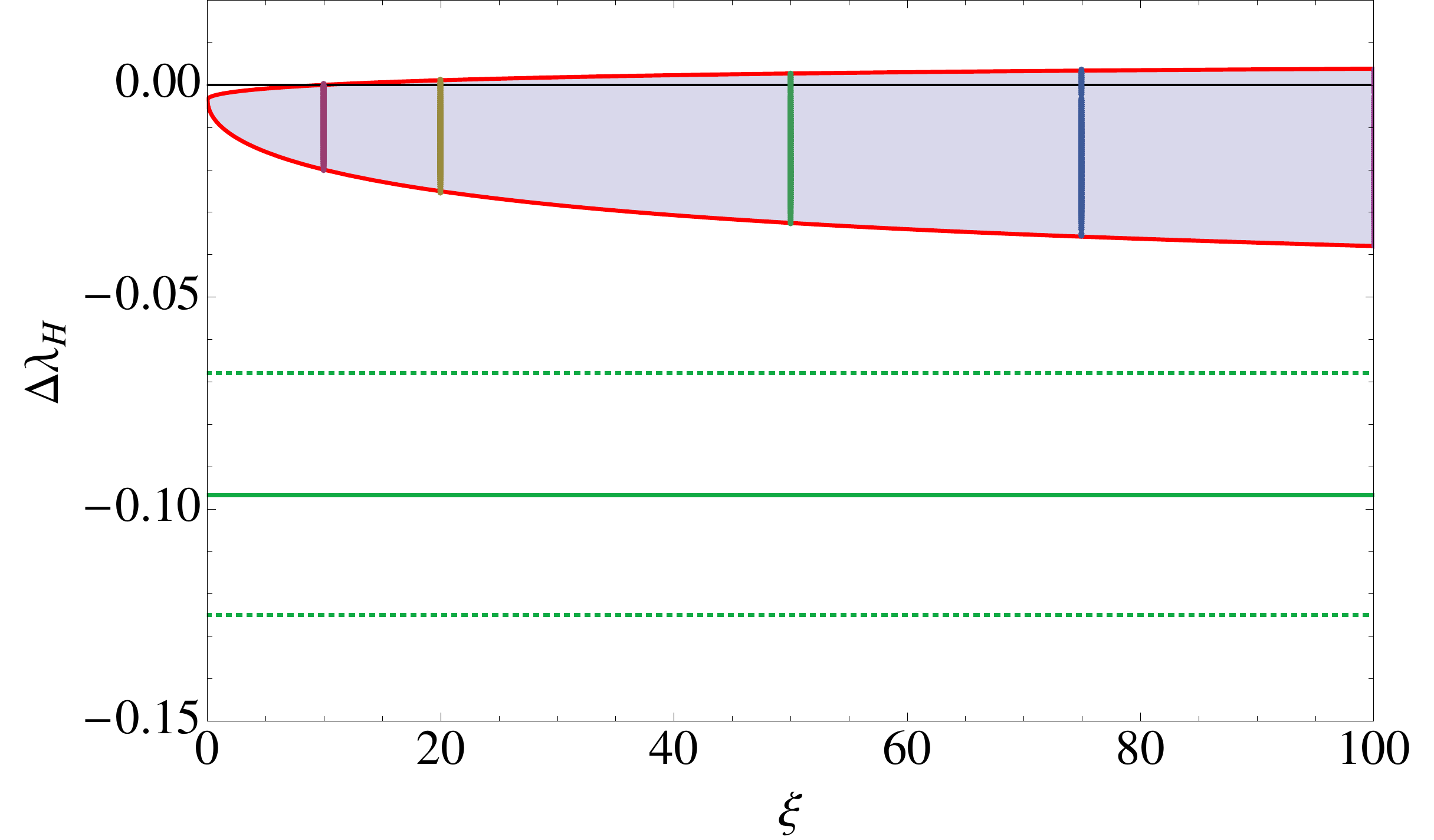}
}
\subfloat[$\tan\beta = 50$]{\includegraphics[width=0.5\textwidth]{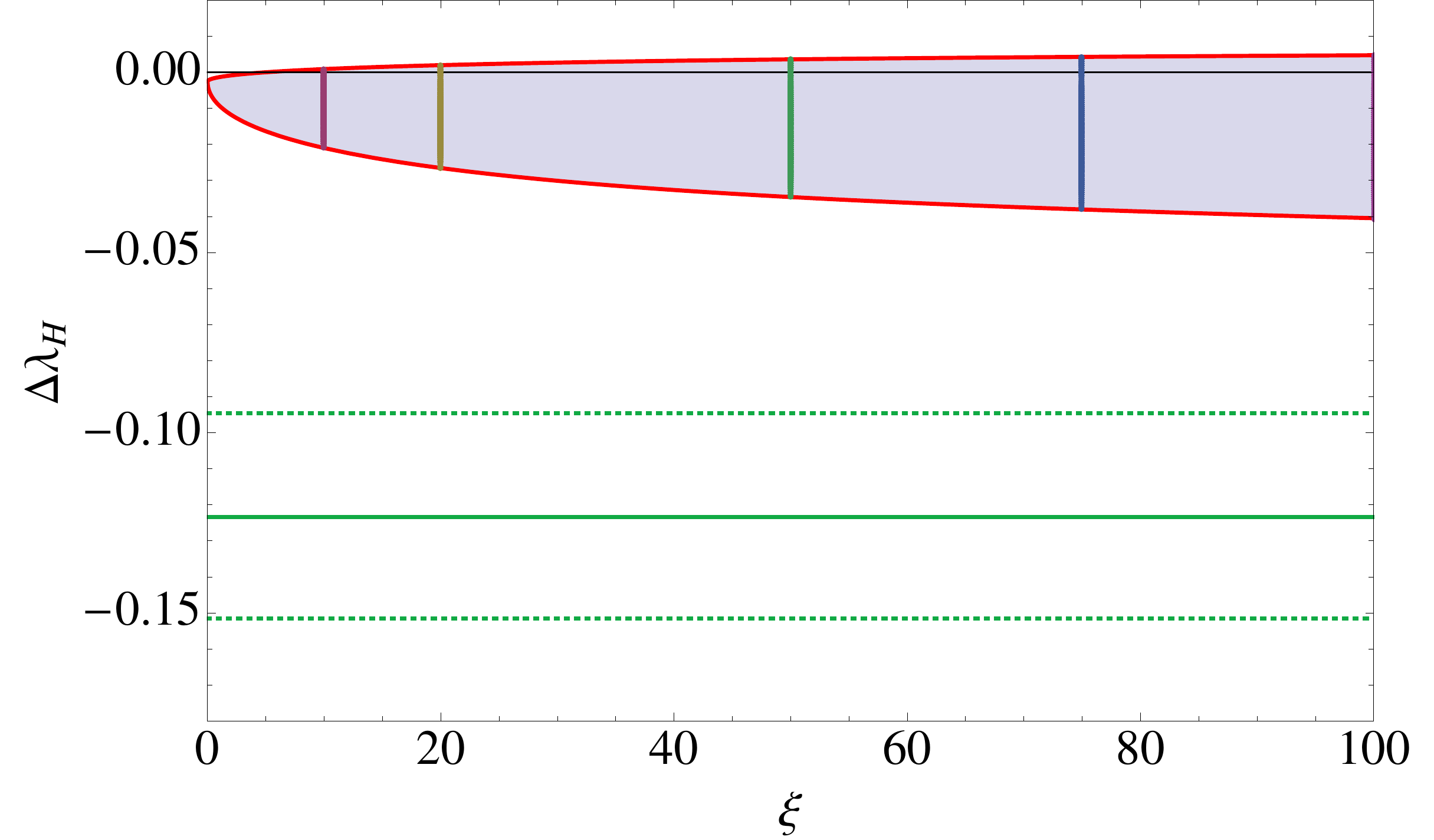}
}
\caption{
These plots show how the required corrections to $\lambda_H$ (solid green) compare to the upper and lower limits of the threshold corrections $\Delta\lambda$ obtained from a scan, at a SUSY scale $\mS = \mStL=\mStR = 10^{10}$ GeV. 
The dashed green lines correspond to the $3\sigma$ upper and lower limits from $m_t$ uncertainty. 
Scanned independently are $ \mu,~ \{ \mSqL, \mSuR, \mSdR, \mSbR, \mSlL, \mSeR, m_A \},~ \{ M_a\}$ and $\tan\beta$. The top trilinear $A_t$ has been set to zero. Each mass is allowed to vary up to $\xi=100$ relative to $\mS$, with scanned points shown here for $\xi=0,10,20,50,75$ and $100$. The four plots show the results for four separate choices of $\tan \beta$.
}

\label{10Vary.FIG}
\end{figure}

\begin{figure}[H]
\centering
\subfloat[$\tan\beta = 1$]{\includegraphics[width=0.5\textwidth]{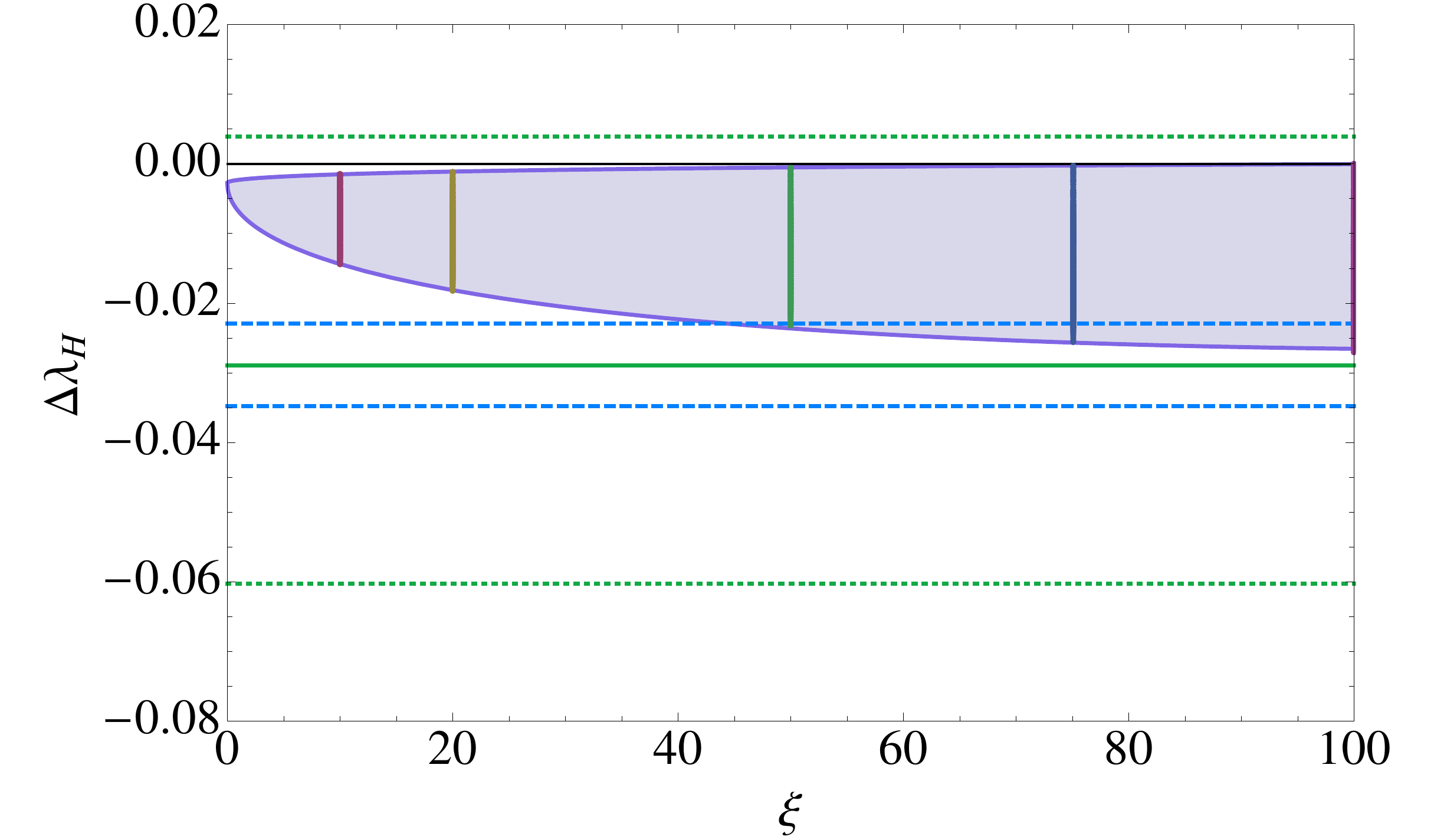}
}
\subfloat[$\tan\beta = 2$]{\includegraphics[width=0.5\textwidth]{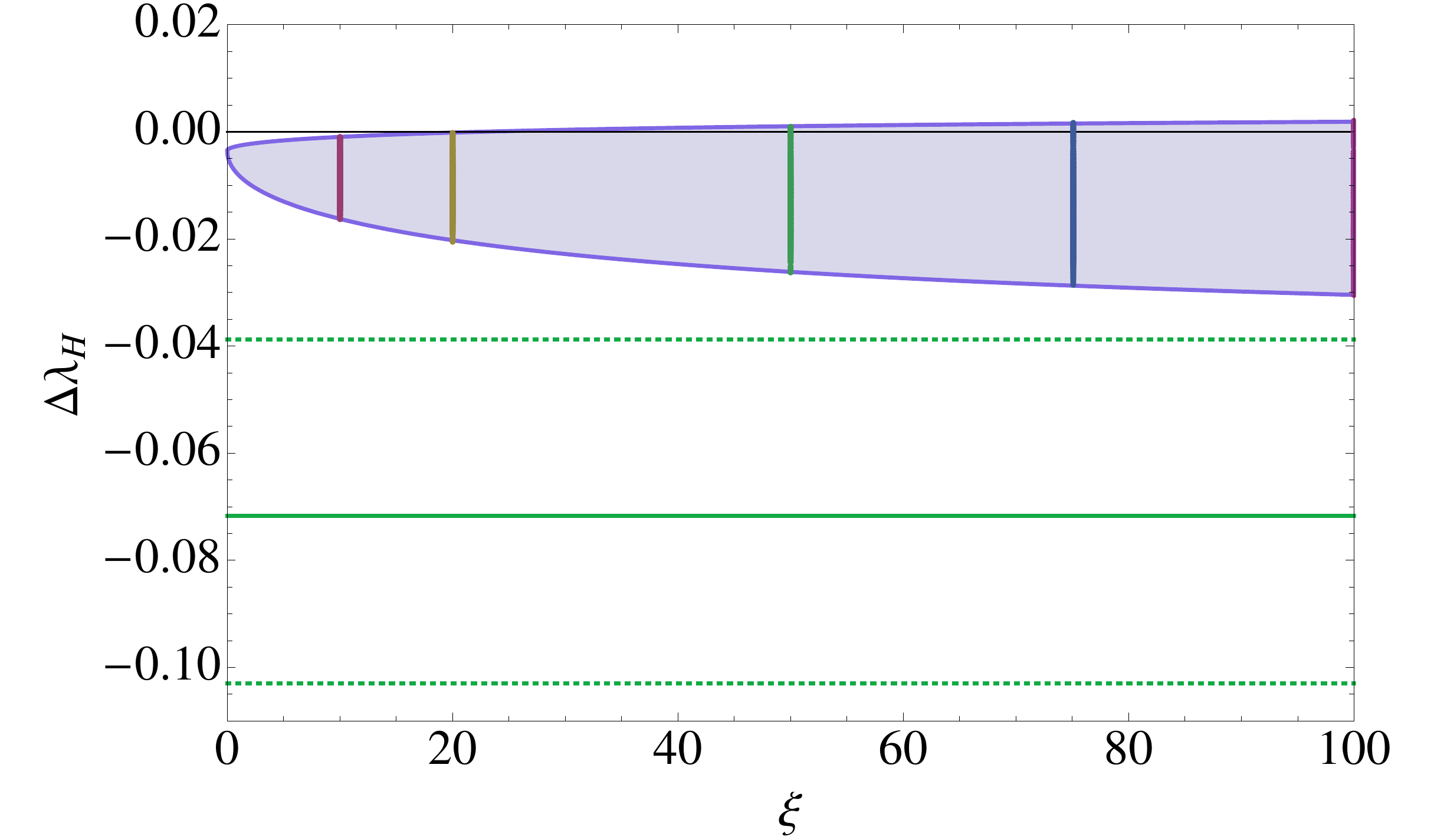}
}
\\
\subfloat[$\tan\beta = 4$]{\includegraphics[width=0.5\textwidth]{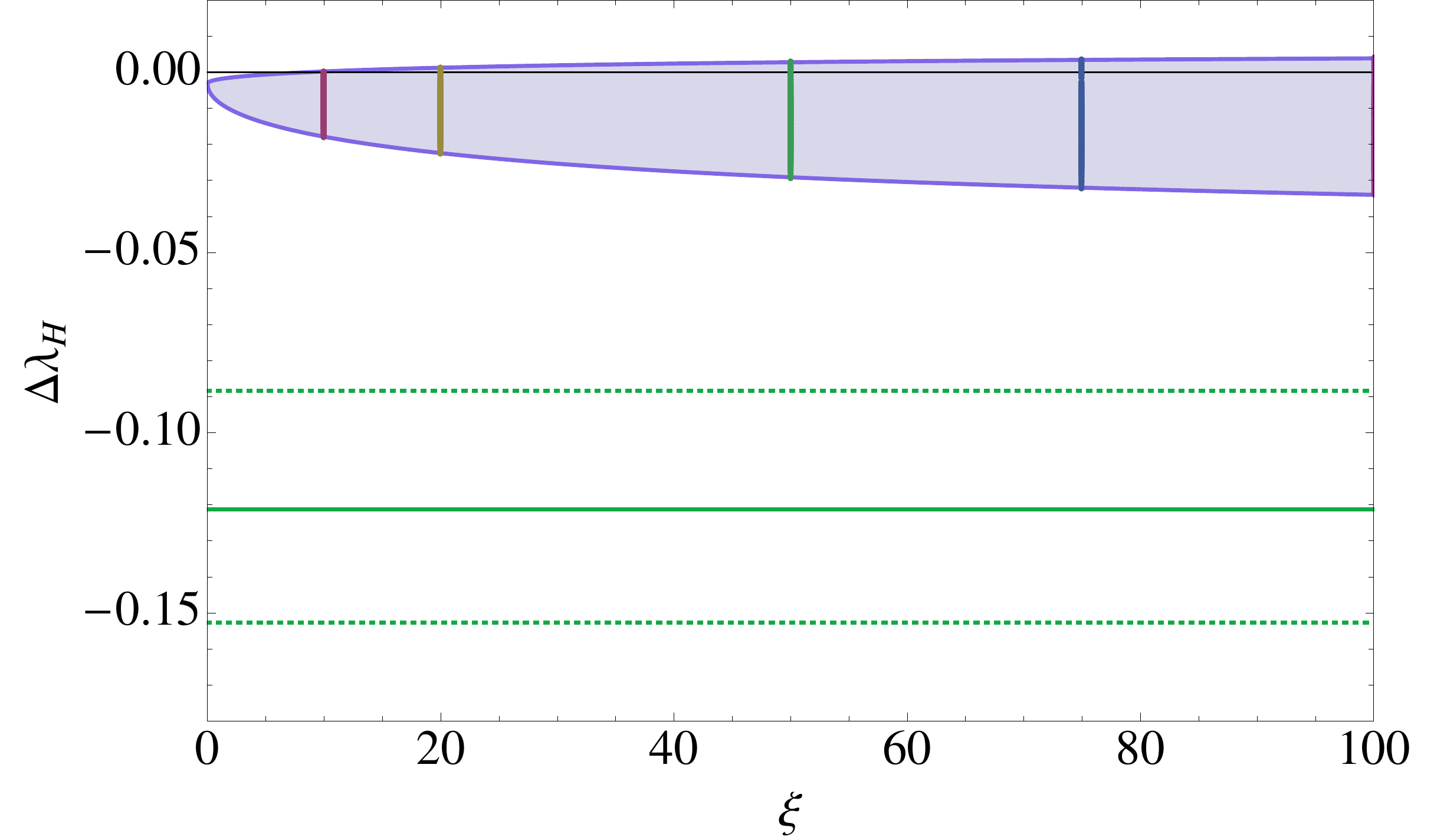}
}
\subfloat[$\tan\beta = 50$]{\includegraphics[width=0.5\textwidth]{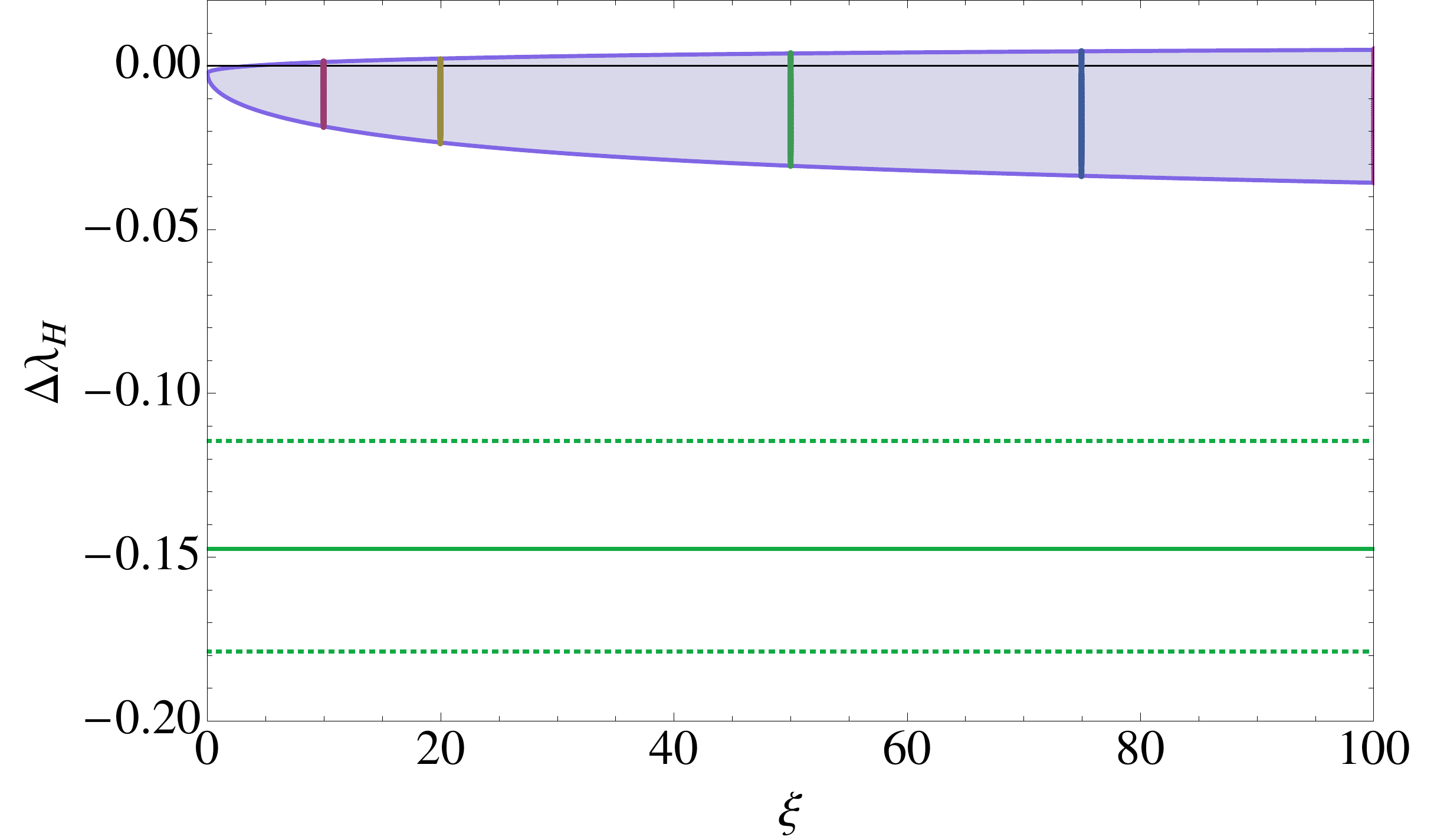}
}
\caption{
These plots show how the required corrections to $\lambda_H$ (solid green) compare to the upper and lower limits of the threshold corrections $\Delta\lambda$ obtained from a scan, at a SUSY scale $\mS = \mStL=\mStR = 10^{16}$ GeV. 
The dashed green lines correspond to the $3\sigma$ upper and lower limits from $m_t$ uncertainty. The dashed blue lines shown for $\tan\beta=1$ correspond to $m_h \pm 1$ GeV.
Scanned independently are $ \mu,~ \{ \mSqL, \mSuR, \mSdR, \mSbR, \mSlL, \mSeR, m_A \},~ \{ M_a\}$ and $\tan\beta$. The top trilinear $A_t$ has been set to zero. Each mass is allowed to vary up to $\xi=100$ relative to $\mS$, with scanned points shown here for $\xi=0,10,20,50,75$ and $100$. The four plots show the results for four separate choices of $\tan \beta$.
}

\label{16Vary.FIG}
\end{figure}

In each of Figs.~\ref{TeVVary.FIG} -- \ref{16Vary.FIG}, we show how $\Delta\lambda^{(1)}$ in the MSSM compares with the required threshold correction for matching to occur (shown by green lines in each figure). We fix the SUSY scale to be $\mS=\mStL=\mStR$, the scale of the LH and RH stops, which we take to be degenerate at the input scale. We then scan over four sets of parameters: \\
\begin{center}
\vspace{-0.25in}
$\{\mSqL, \mSuR, \mSdR,\mSbR, \mSlL, \mSeR, m_A \}$, $\{\mu \}$, $\{M_a\}$, $\{\tan\beta\}$\ ,
\vspace{0.1in}
\end{center}
where $M_a$ is the gaugino mass parameter. We assume the gaugino masses obey the standard GUT relation, namely

\beq
\frac{M_1}{g_1^2} = \frac{M_2}{g_2^2} = \frac{M_3}{g_3^2} \ ,
\label{GUTrelation.EQ}
\eeq 
so that when we vary $M_a$, we choose it to be equivalent to the Wino mass parameter $M_2$, with $M_1$ related to $M_a$ by the above expression.

We allow the scalar and gaugino mass parameters above to vary in the range $\mS \leq \{\mSqL, \mSuR, \mSdR,\mSbR, \mSlL, \mSeR, m_A \},\{M_a \} \leq 100\ \mS$ , and the higgsino mass parameter $\mu$ to vary from  $\mS /  100 \leq \mu \leq \mS$, and investigate various choices of $\tan\beta$. We do not allow the higgsino mass $\mu$ to vary above $\mS$ so as to not run afoul of stop mixing constraints.

We study the values of $\tan\beta=$ 1, 2, 4 and 50 initially, with further fine-graining as necessary to determine the exact range where matching can be achieved at a given $\mS$. The gluino mass $M_3$ only appears in the two-loop threshold corrections to $\lambda_H$ \cite{Bagnaschi:2014rsa}, and therefore is not a parameter we vary explicitly. 

We define the parameter $\xi$ used in Figs.~\ref{TeVVary.FIG} -- \ref{16Vary.FIG} as follows. For any value of $\xi$, the superpartner masses in the MSSM spectrum are allowed to vary either from $\mS$ to as large as $\xi \ \mS$ for the scalars and gauginos, or from $\mS$ to as small as $\mS / \xi$ for the higgsinos, where $\mS$ is the scale set by the degenerate stop masses $m_{\tilde q_{L,3}}$ and $m_{\tilde t_{R,3}}$. For example, in Fig.~\ref{TeVVary.FIG} if the $x$-axis $\xi=40$, the superpartner masses are allowed to vary from $\mS$ to as low as $\mS/40$ for higgsinos, and as high as $40\mS$ for scalars and gauginos.

Now we come to the interpretation of Figs. ~\ref{TeVVary.FIG} -- \ref{16Vary.FIG}. As expected, matching the SM quartic coupling to the SUSY theory at one-loop for $\mS=5\times10^3$ GeV is straightforward. Having TeV scale SUSY ``prefers" larger values of $\tan\beta$, but is compatible with a wide range of values of $\tan\beta$ depending on the variation of the superpartner masses. It also prefers having a relatively large trilinear $A_t$. As we go to higher superpartner mass scales one finds that significant one-loop threshold corrections are required in order for matching with the SM value to occur, which in turn can only be achieved by large variations or non-degeneracies in the superpartner masses, thereby creating large  logarithms that saturate the matching condition requirement. 

For example, let us look at the $\tan\beta=2$ plot for the $\mS=10^{10}\gev$ case (Fig.~\ref{10Vary.FIG}b). We see that if there are no variations or small variations (under a factor of 10) among the superpartner masses near $10^{10}\gev$ there is no way to achieve large enough threshold corrections to match to the necessary SM Higgs self coupling at the scale $\mS$. One needs variations greater than a factor of 10 to match the $3\sigma_{m_t}$ upper limit on $\lambda_H$. Significantly larger non-degeneracy would be required to match the central value of $\lambda_H$. These variations are substantially larger than typically considered in supersymmetry breaking schemes. 
One could only consider such large variations in the context of an underlying supersymmetry-breaking scheme that naturally gave rise to large non-degeneracies. The issue then becomes highly model-dependent, so we do not discuss it further here.

In \cite{Vega:2015fna}, it was pointed out that at very large values of $\tan\beta \sim 200$, the sbottom and stau mixing alone are enough for a degenerate SUSY spectrum at $\mS \sim 10^{16}\ \gev$ to give the correct Higgs mass. However, the value of the superpotential bottom Yukawa coupling $\hat{y}_b = y_b /  \cos\beta$ is such that $\hat{\alpha}_b = \hat{y}_b^2 / 4\pi$ is so large that there is a Landau pole at $\Lambda \sim 10 \mS$ \cite{Vega:2015fna}. In our analysis we have shown that the low $\tan\beta \sim 1$ regime also allows the correct Higgs boson mass to be obtained within $1\sigma$ if there is non-degeneracy $\xi \gtrsim10$, and within 1 GeV if there is non-degeneracy $\xi \gtrsim 45$, as seen in Fig.\ \ref{16Vary.FIG}, while avoiding the potentially dangerous effects of having  large $\tan\beta$. If one allows for even larger non-degeneracy, $\xi \gtrsim 100$, the central value for the Higgs mass is reached.

We remind the reader that while at tree-level, the mass of the physical propagating Higgs boson in the Standard Model is $m_h^2 \propto \lambda_H$, there are uncertainties due to the top quark mass and the QCD coupling which are not insignificant (see for example \cite{Buttazzo:2013uya, Martin:2014cxa}). Therefore, only with better understanding of the exact relationship between the Higgs boson mass and $\lambda_H$ in the Standard Model will we be able to precisely determine the ability to match on to a SUSY theory. Additionally, there are theoretical uncertainties associated with missing higher order corrections to the matching of SUSY at high scales, so that overall, one expects a typical $\Delta m_h^{th.} \sim 1$ GeV for high-scale SUSY \cite{Vega:2015fna}.

\section{Example Spectra}
\label{Examples.SEC}
\bigskip

In this section, we consider various benchmark spectra which allow for matching $\lambda_H^{SM}$ to $\lambda_H^{tree}+\Delta\lambda_H$ at the chosen SUSY scales $\mS= 5\times10^3,~10^6,~10^{10}$ and $10^{16}$ GeV. We choose the benchmark spectra by a $\chi^2$-minimisation, with the $\chi^2$ being obtained as
\begin{align}
\chi^2 =\left( \frac{\Delta\lambda_H^{obt.} - \Delta\lambda_H^{req.}}{\sigma_{\Delta\lambda_H}}\right)^2 \ ,
\label{chiSq.EQ}
\end{align}
where the error $\sigma_{\Delta\lambda_H}$ is obtained by using the 1-$\sigma$ error associated with the top mass $m_t(m_t)$. The quantities $\Delta\lambda_H^{obt.}$ and $\Delta\lambda_H^{req.}$ are the obtained and required threshold corrections at one loop from the SUSY spectrum respectively. For all the example spectra in this section, we have cross-checked with the public code \texttt{SusyHD} \cite{Vega:2015fna} that our calculation agrees with theirs.

For TeV-scale SUSY with $\tan\beta=50$, we find that a spectrum where all the scalars are degenerate, $m_{scalars}=\mS =5\times10^3$ GeV, the Gaugino masses are $M_a = 9\mS$, and $\mu = 0.5 \mS$ has $\chi^2=3\times10^{-4}$, so that $\Delta\lambda_H^{req.}$ and $\Delta\lambda_H^{obt.}$ are in good agreement. To ensure that there is no undue effect of the choice of matching scale, we can examine how the various components of $\lambda_H$ and $\Delta\lambda_H$ conspire to match the Standard Model $\lambda_H^{SM}$, as a function of the matching scale. This is shown in Fig. \ref{TeV_matching.FIG} below. We see that the choice of $\tan\beta=50$ requires some positive threshold corrections, since $\lambda_H^{SM}$ is greater than $\lambda_H^{tree}$. This is obtained with a fairly large positive correction from stop-mixing with just a small negative threshold correction being provided primarily by the Higgsinos/gauginos (hence the non-degeneracy of the Higgsino with the gauginos). The correction from the other scalars
is small for this choice of parameters.

\begin{figure}[t]
\centering
\includegraphics[scale=0.9]{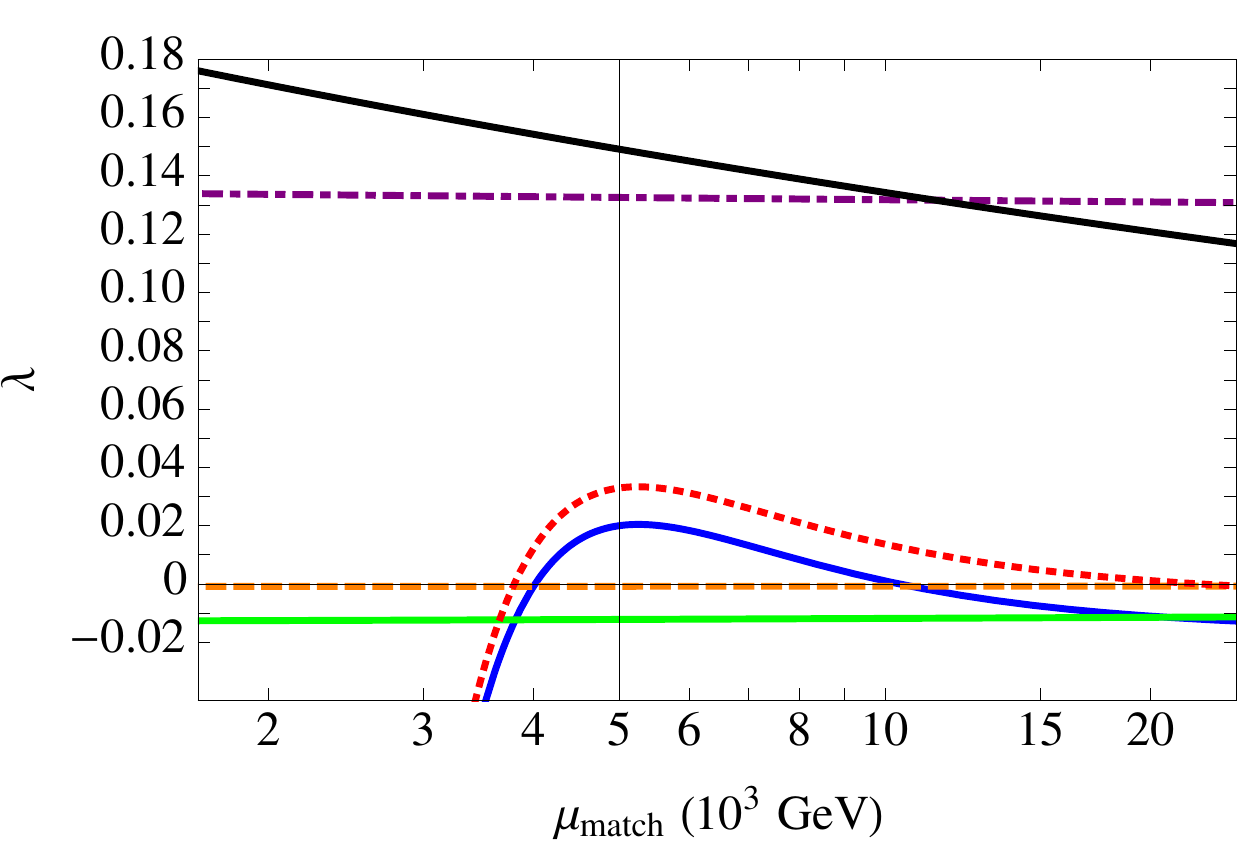}
\caption{Shown are $\lambda_H^{SM}$ (black), $\lambda_H^{tree}$ (purple dot-dashed) and $\Delta\lambda_H^{tot.}$ (blue), as a function of the matching scale $\mu_{match}= \mStL =\mStR$. The benchmark point was obtained for matching at $\mu_{match}=5\times10^3$ GeV. Also shown are the various components that add up to give $\Delta\lambda_H^{tot.}$, namely the correction from $\MS$ to $\DR$ (orange dashed), the scalar contribution (red dotted) and the Higgsino/gaugino contribution (green solid).}
\label{TeV_matching.FIG}
\end{figure}

For PeV-scale SUSY with $\tan\beta=2.5$, we find that a spectrum where all the scalars are degenerate, $m_{scalars}=\mS$, $M_a = \mu = \mS$, has $\chi^2=0.004$, so that $\Delta\lambda_H^{req.}$ and $\Delta\lambda_H^{obt.}$ are in good agreement. Again, to ensure that there is no undue effect of the choice of matching scale, we can examine how the various components of $\lambda_H$ and $\Delta\lambda_H$ conspire to match the Standard Model $\lambda_H^{SM}$, as a function of the matching scale. This is shown in Fig. \ref{PeV_matching.FIG} below. We see that the choice of $\tan\beta=2.5$ requires threshold corrections of order ($-\lambda^{tree}_H/20$), which are provided primarily by the Higgsinos/gauginos. The correction from scalars, which includes the $L\leftrightarrow R$ mixing of the stops is almost zero for this choice of parameters, therefore our choice of setting $A_t=0$ does not affect the result. We note that for matching scales going below $\mu_{match}= \mStL =\mStR=10^6$ GeV, the effect of the scalars (red dotted) increases. This is the well-known threshold correction due to the stop mixing parameter $X_t = A_t - \mu \cot\beta$ that enters in the last three lines in Eq. (\ref{ThirdGenPart.EQ}). In that range, the impact of the choice of $A_t$ is important, and can be used to cancel off the effect of $\mu\cot\beta > \mStL,\mStR$.

\begin{figure}[t]
\centering
\includegraphics[scale=0.9]{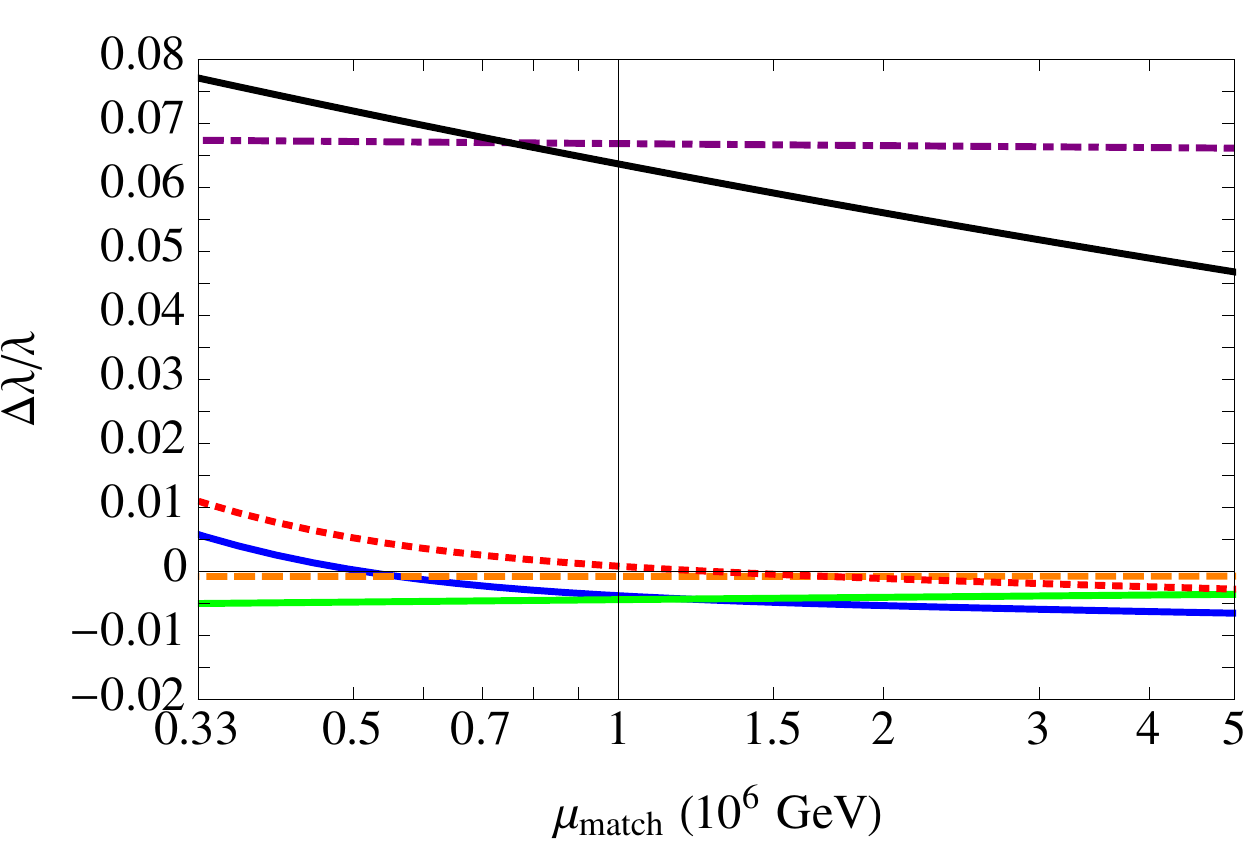}
\caption{Shown are $\lambda_H^{SM}$ (black), $\lambda_H^{tree}$ (purple dot-dashed) and $\Delta\lambda_H^{tot.}$ (blue), as a function of the matching scale $\mu_{match} = \mStL =\mStR$. The benchmark point was obtained for matching at $\mu_{match}=10^6$ GeV. Also shown are the various components that add up to give $\Delta\lambda_H^{tot.}$, namely the correction from $\MS$ to $\DR$ (orange dashed), the scalar contribution (red dotted) and the Higgsino/gaugino contribution (green solid).}
\label{PeV_matching.FIG}
\end{figure}

For an intermediate SUSY scale of $10^{10}$ GeV with $\tan\beta=1$, we find that a spectrum where all the superpartners are degenerate, $m_{scalars}=M_a =\mu=\mS$, has $\chi^2=0.008$, so that $\Delta\lambda_H^{req.}$ and $\Delta\lambda_H^{obt.}$ are in good agreement. Once more, to ensure that there is no undue effect of the choice of matching scale, we can examine how the various components of $\lambda_H$ and $\Delta\lambda_H$ conspire to match the Standard Model $\lambda_H^{SM}$, as a function of the matching scale. This is shown in Fig. \ref{10_matching.FIG} below. We see that the choice of $\tan\beta=1$ requires threshold corrections to be of order $-\lambda^{SM}_H$, since the tree-level value of $\lambda_H^{tree}=0$. Here, the threshold corrections are given by a partial cancellation between the positive corrections due to the scalars and a negative correction due to the Higgsinos/Gauginos, with a small contribution from the change of renormalisation scheme. This cancellation effect at $\tan\beta \sim 1$ can be understood by referring back to Fig. \ref{DeltaLambdaTB.FIG}, where we see that for $A_t = 0$, the stop mixing and the higgsino/gaugino contributions to the threshold corrections are similar in size, and opposite in sign. We note that for matching scales going below $\mu_{match}= \mStL =\mStR=10^{10}$ GeV, the effect of the scalars (red dotted) increases. This is the well-known threshold correction due to the stop mixing parameter $X_t = A_t - \mu \cot\beta$ that enters in the last three lines in Eq. (\ref{ThirdGenPart.EQ}). In that range, the impact of the choice of $A_t$ is important, and can be used to cancel off the effect of $\mu\cot\beta > \mStL,\mStR$.

\begin{figure}[t]
\centering
\includegraphics[scale=0.9]{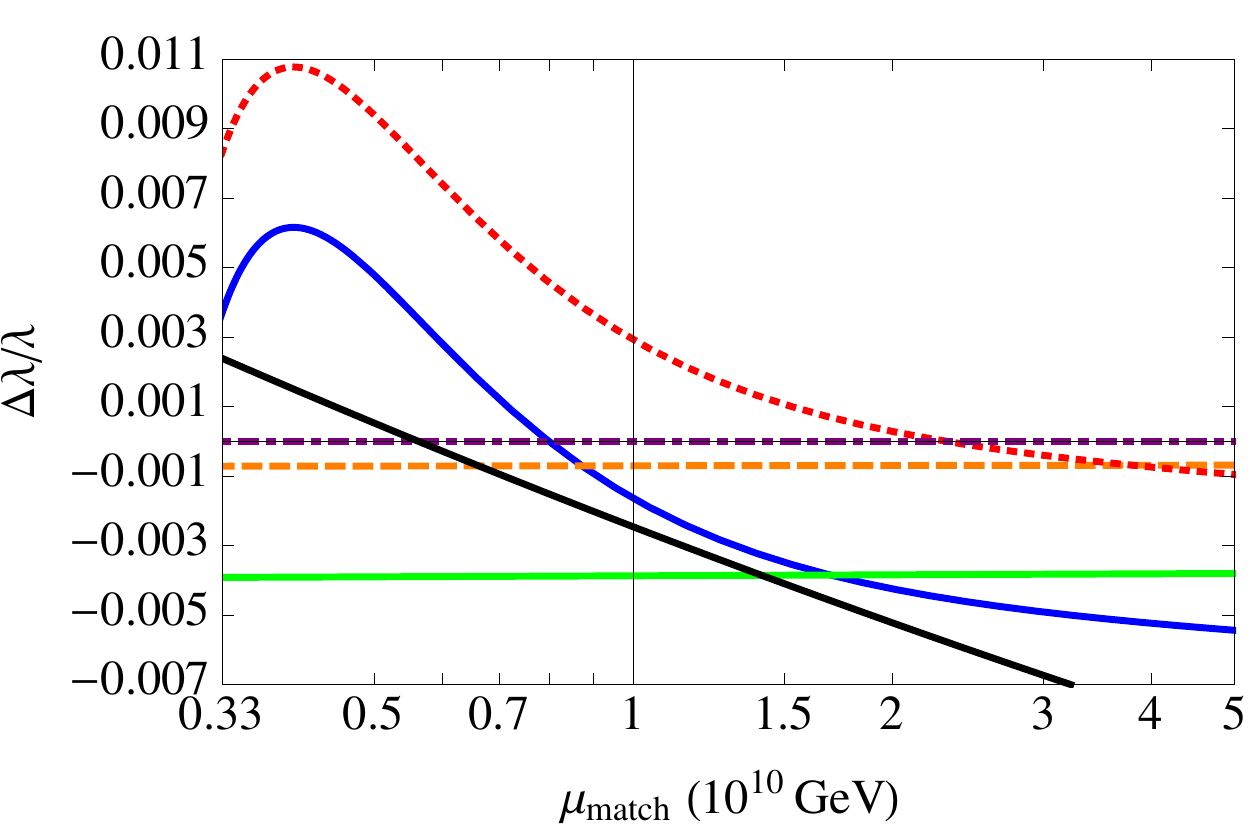}
\caption{Shown are $\lambda_H^{SM}$ (black), $\lambda_H^{tree}$ (purple dot-dashed) and $\Delta\lambda_H^{tot.}$ (blue), as a function of the matching scale $\mu_{match} = \mStL =\mStR$. The benchmark point was obtained for matching at $\mu_{match}=10^{10}$ GeV. Also shown are the various components that add up to give $\Delta\lambda_H^{tot.}$, namely the correction from $\MS$ to $\DR$ (orange dashed), the scalar contribution (red dotted) and the Higgsino/gaugino contribution (green solid).}
\label{10_matching.FIG}
\end{figure}

For SUSY at the GUT scale of $10^{16}$ GeV with $\tan\beta=1$, we find that a spectrum where the superpartner masses are given by $m_{scalars}=\mS$, $M_a =20\mS$ and $\mu = 0.001 \mS$, has $\chi^2=0.0001$, so that $\Delta\lambda_H^{req.}$ and $\Delta\lambda_H^{obt.}$ are in good agreement. Once more, to ensure that there is no undue effect of the choice of matching scale, we can examine how the various components of $\lambda_H$ and $\Delta\lambda_H$ conspire to match the Standard Model $\lambda_H^{SM}$, as a function of the matching scale. This is shown in Fig. \ref{16_matching.FIG} below. We see that the choice of $\tan\beta=1$ requires threshold corrections to be $-\lambda^{SM}_H$, since the tree-level value of $\lambda_H^{tree}=0$. Here, the threshold corrections are almost entirely due to the negative correction from the Higgsinos/Gauginos, with a small contribution from the change of renormalisation scheme. Since the stop-mixing contribution is almost zero, the choice of $A_t=0$ does not impact our results.

\begin{figure}[t]
\centering
\includegraphics[scale=0.9]{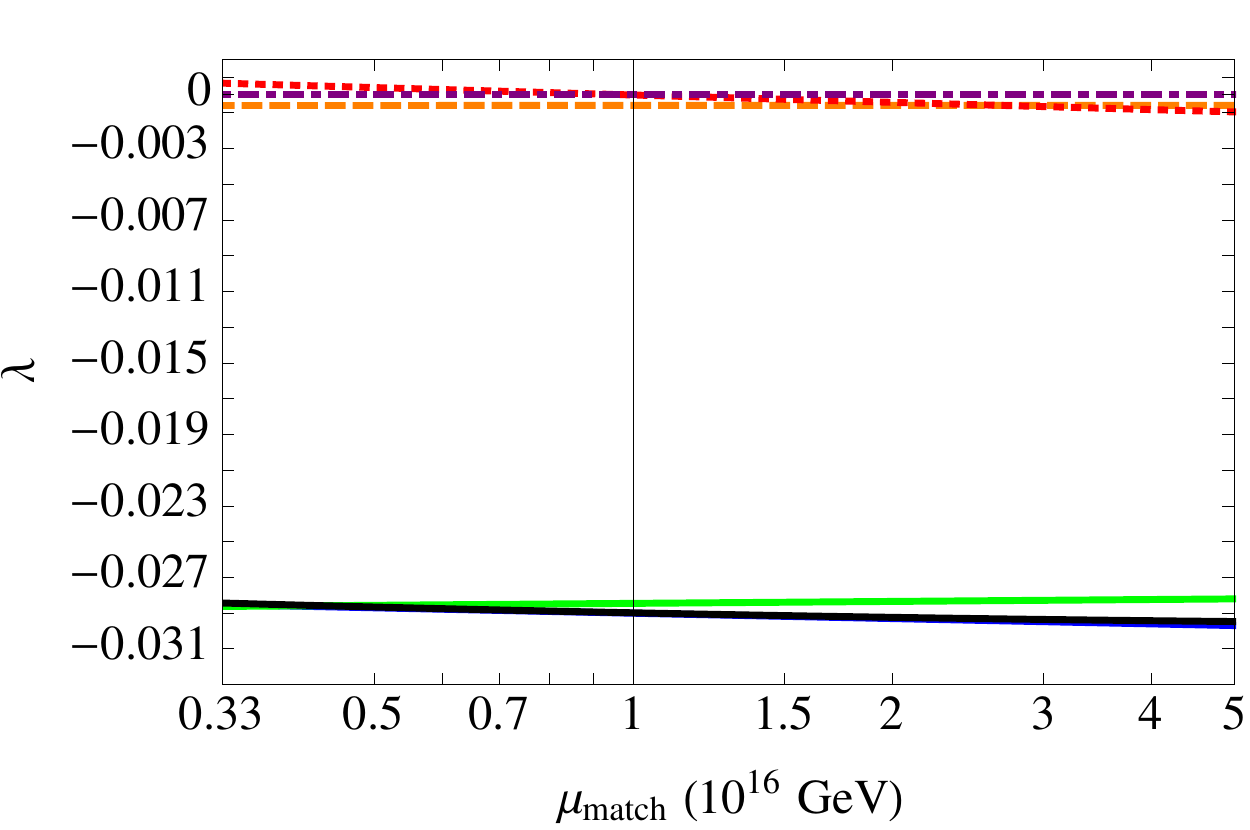}
\caption{Shown are $\lambda_H^{SM}$ (black), $\lambda_H^{tree}$ (purple dot-dashed) and $\Delta\lambda_H^{tot.}$ (blue), as a function of the matching scale $\mu_{match} = \mStL =\mStR$. The benchmark point was obtained for matching at $\mu_{match}=10^{16}$ GeV. Also shown are the various components that add up to give $\Delta\lambda_H^{tot.}$, namely the correction from $\MS$ to $\DR$ (orange dashed), the scalar contribution (red dotted) and the Higgsino/gaugino contribution (green solid).}
\label{16_matching.FIG}
\end{figure}

To conclude this subsection, we show two-dimensional plots where the spectra match $\lambda^{SM}_H = \lambda^{tree}_H + \Delta\lambda_H$ to within 1, 1.96 and 3 sigma (green, orange, red), where sigma is the same $\sigma_{\Delta\lambda_H}$ as in Eq. (\ref{chiSq.EQ}). We choose to plot the gaugino mass $M_a=M_2$ against the Higgsino mass parameter $\mu$, since the threshold corrections depend strongly on the choice of these two parameters. We set $M_1 = (g_1^2/g_2^2) M_2$ as in Eq. (\ref{GUTrelation.EQ}), so as to maintain the GUT relation between gaugino masses. Showing such contours for $\mS=5\times10^3$ is not particularly illuminating, since there are many regions in parameter space that can have matching of $\lambda_H$. However, for the higher SUSY scales, it is interesting to consider what regions are viable, and how they shift as a function of the choice of matching scale.

\begin{figure}[t]
\centering
\subfloat[tan$\beta=2.5$]{
\includegraphics[width=0.49\textwidth]{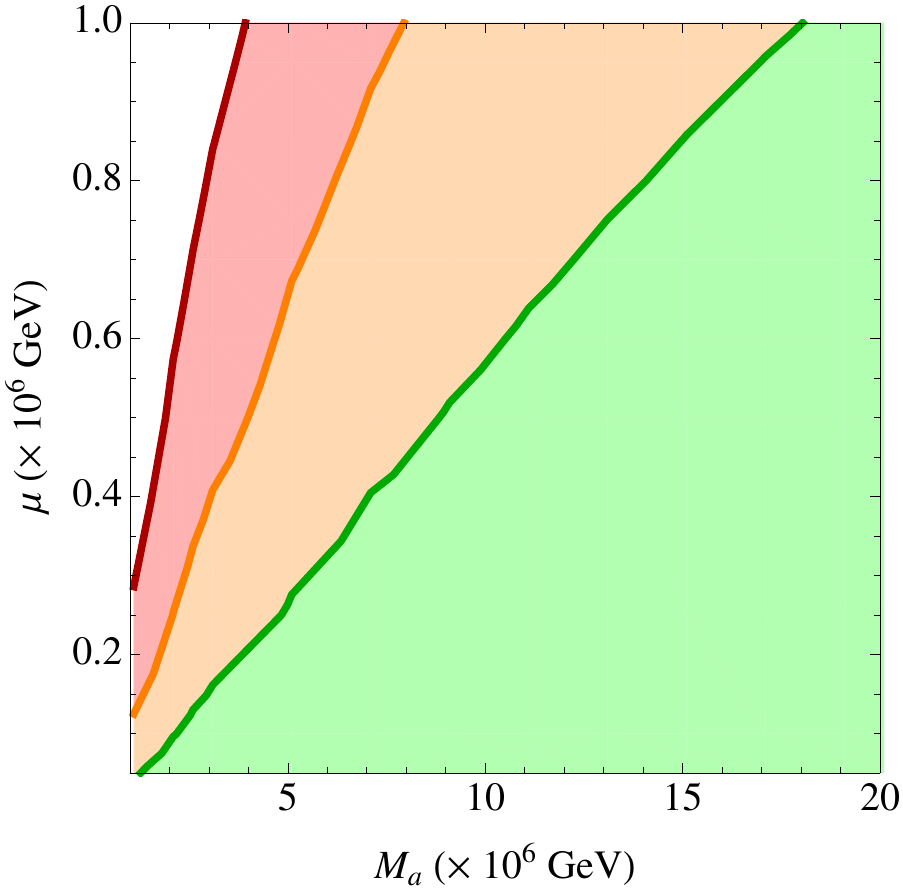}} ~~ \subfloat[tan$\beta=3.1$]{
\includegraphics[width=0.49\textwidth]{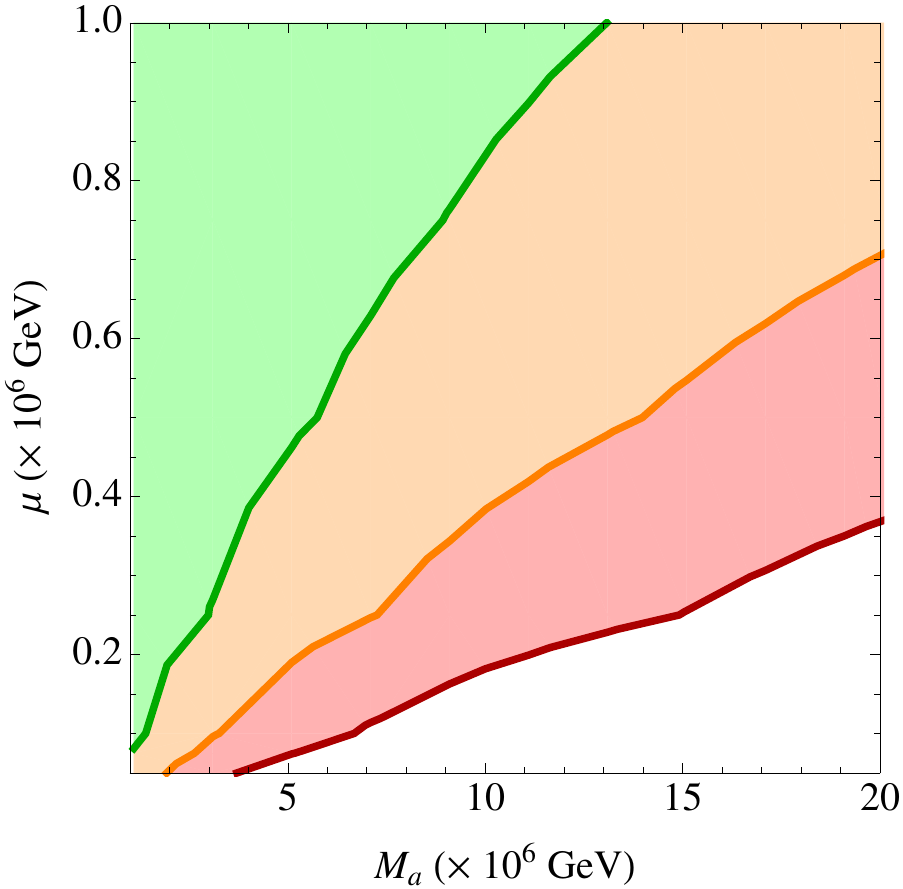}} 
\caption{Shown are the regions in $M_a, \mu$ space, for degenerate scalars $\mS$, $\tan\beta=2.5,~3.1$ and $\mStL=\mStR=\mu_{match}$ where $\lambda_H$ is matched to within 1, 1.96 and 3 sigma (green, orange, red). The matching has been performed at the scale $\mu_{match}=10^{6}$ GeV.
}
\label{PeV_contours.FIG}
\end{figure}

We see in Fig. \ref{PeV_contours.FIG} that for PeV-scale SUSY, shifting only slightly the value of $\tan\beta$ results in a large change in the ratio of $\mu/M_a$ that is necessary for matching to occur. For $\tan\beta=3.1$, degeneracy, or only slight non-degeneracy is required, with ratios $\Order(1)$ viable, while for $\tan\beta=2.5$, ratios must be $\Order(10)$ or greater. This can be better understood by referring back to Fig. \ref{DeltaLambdaPlotTB.FIG}, which gave an idea of the ``optimal" values of $\tan\beta$ for matching, for various different SUSY scales. We see that regardless of the SUSY scale, $\Delta\lambda_H^{req.}$ has a strong $\tan\beta$ dependence between $1 \leq \tan\beta \lesssim 5$, with the strongest dependence near $\tan\beta\sim2-3$. Since for PeV-scale SUSY, the optimal $\tan\beta \sim 3$, we can see that small variations in $\tan\beta$ around that value will result in large changes in $\Delta\lambda_H^{req.}$, thus requiring a large change in the ratio $\mu/M_a$.

\begin{figure}[t]
\centering
\subfloat[tan$\beta=1$]{
\includegraphics[width=0.49\textwidth]{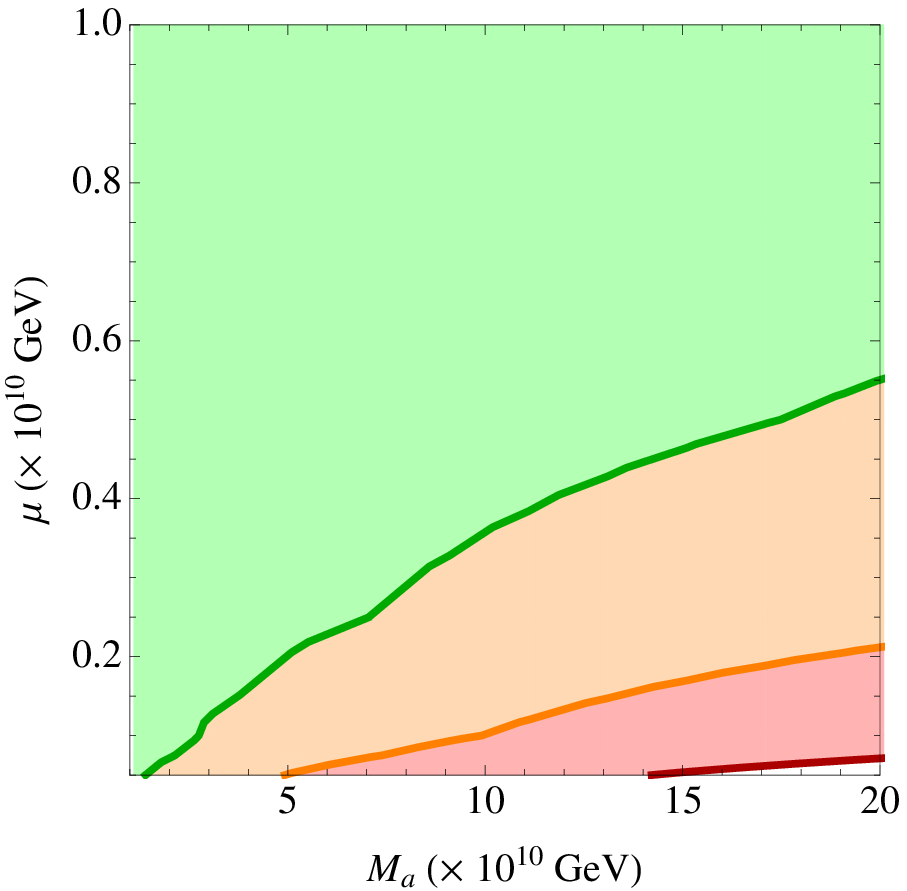}} ~~ \subfloat[tan$\beta=1.2$]{
\includegraphics[width=0.49\textwidth]{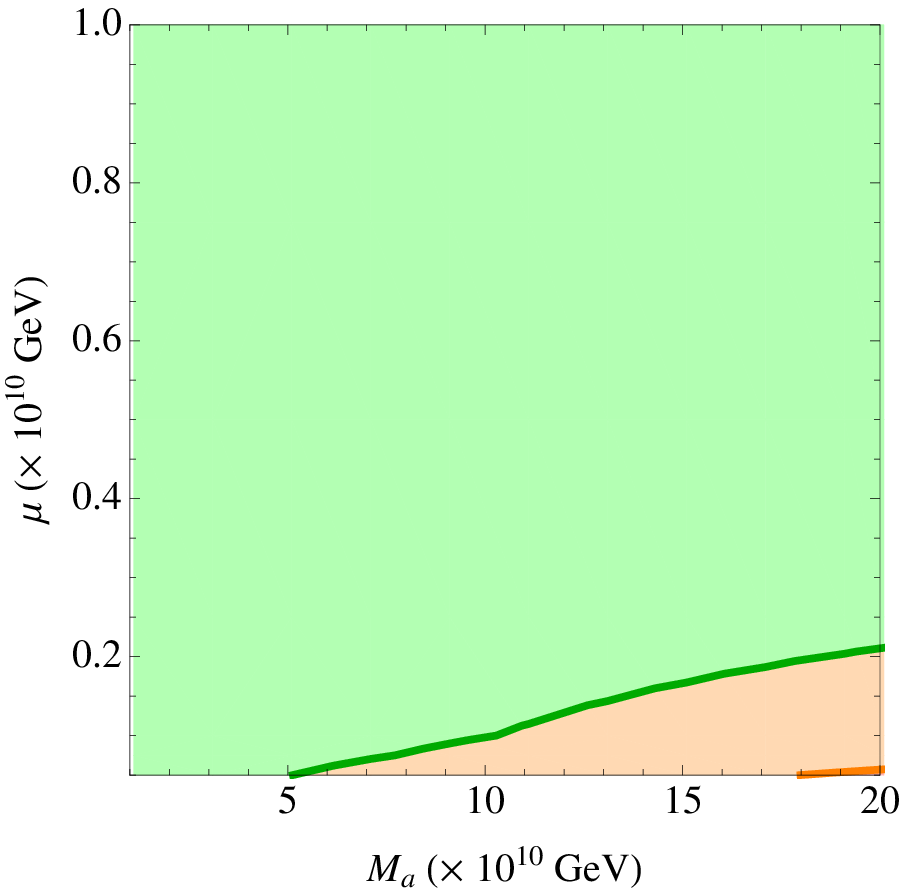}} 
\caption{Shown are the regions in $M_a, \mu$ space, for degenerate scalars $\mS$, $\tan\beta=1,~1.2$ and $\mStL=\mStR=\mu_{match}$ where $\lambda_H$ is matched to within 1, 1.96 and 3 sigma (green, orange, red). The matching has been performed at the scale $\mu_{match}=10^{10}$ GeV.
}
\label{10_contours.FIG}
\end{figure}

For SUSY at $\mS=10^{10}$ GeV, we see in Fig. \ref{10_contours.FIG} that for both $\tan\beta=1$ and $\tan\beta=1.2$, ratios of $\mu/M_a \sim 1$ are sufficient to achieve matching. This makes sense, since $\mu = 10^{10}$ GeV is close to the scale where $\lambda_H^{SM}$ passes through zero, which would therefore match with the tree-level value of $\lambda_H^{tree}(\tan\beta=1) = 0$. Values away from $\tan\beta=1$ actually expand the parameter space available for matching to occur. This is because for $\tan\beta=1$, some of the contributions to $\Delta\lambda^{(1)}_H$ are zero, as can be seen in Fig. \ref{DeltaLambdaTB.FIG}, and Eqs. (\ref{SchemeSwitch.EQ}--\ref{Gauginos1L.EQ}). By going to values $\tan\beta \neq 1$, contributions which were previously zero can now be used to open up more regions of the parameter space.

\begin{figure}[t]
\centering
\subfloat[tan$\beta=0.9$]{
\includegraphics[width=0.49\textwidth]{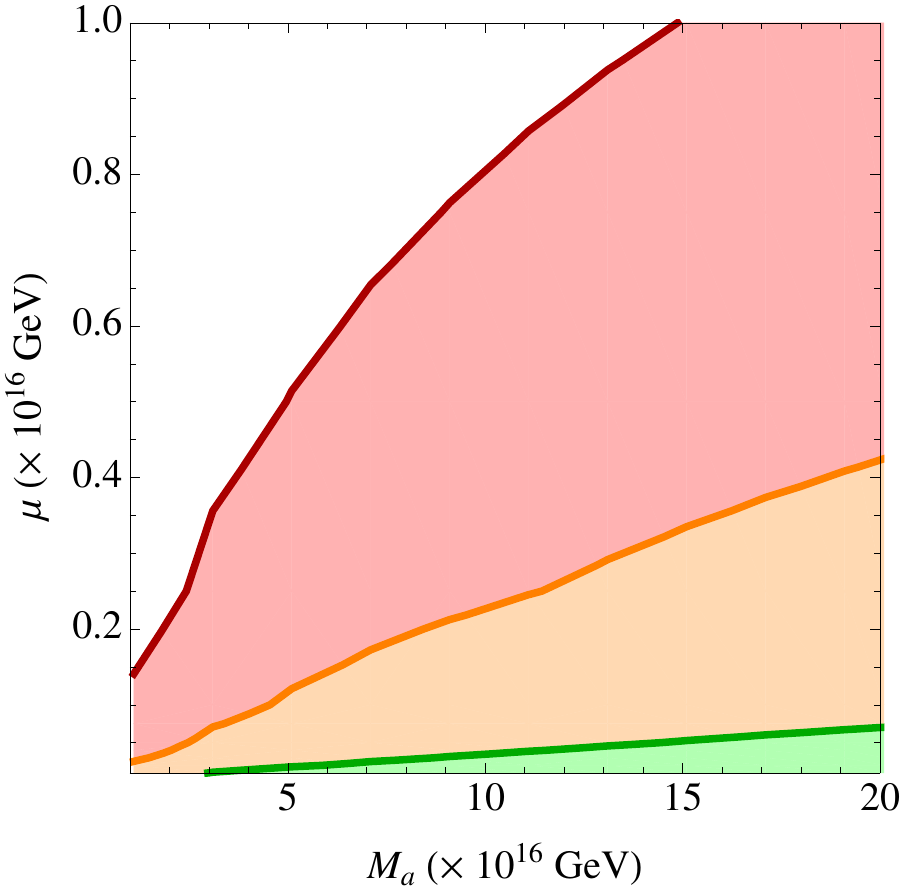}}~~
\subfloat[tan$\beta=1$]{
\includegraphics[width=0.49\textwidth]{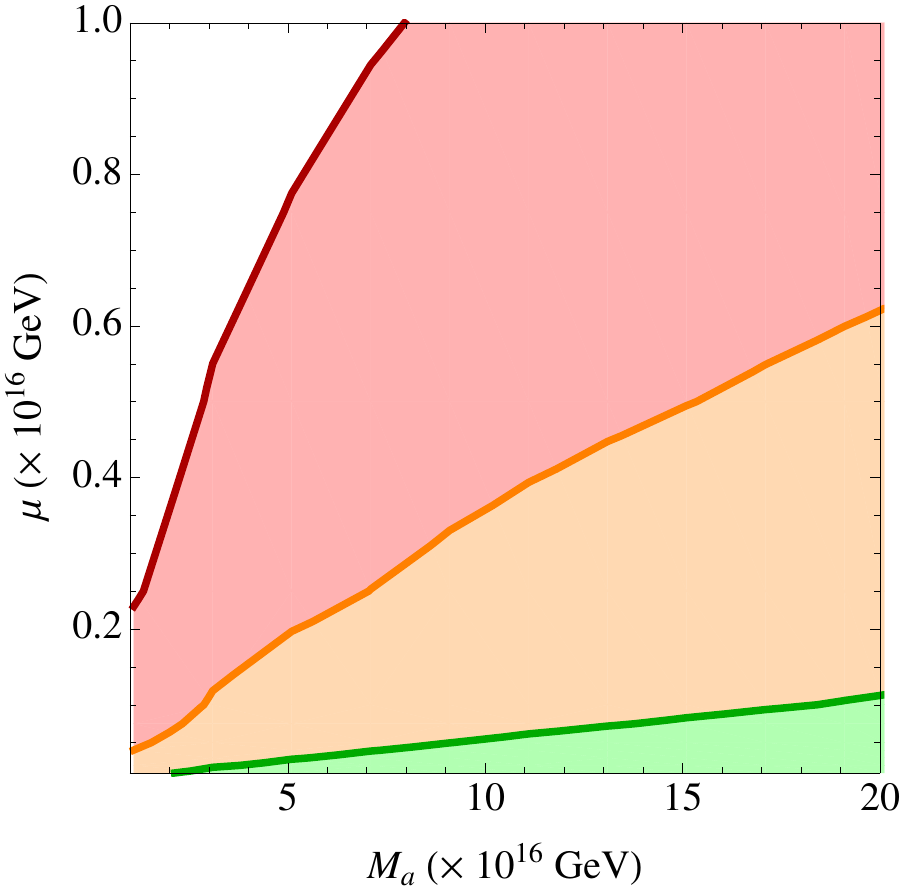}} 
\caption{Shown are the regions in $M_a, \mu$ space, for degenerate scalars $\mS$, $\tan\beta=1,~0.9$ and $\mStL=\mStR=\mu_{match}$ where $\lambda_H$ is matched to within 1, 1.96 and 3 sigma (green, orange, red). The matching has been performed at the scale $\mu_{match}=10^{16}$ GeV.
}
\label{16_contours.FIG}
\end{figure}

From Fig. \ref{16_contours.FIG}, we see that for GUT-scale SUSY at $10^{16}$ GeV, the ratio $\mu / M_a$ has to be of $\Order(50)$ or greater. This is not an unreasonably large non-degeneracy, because the two mass parameters are typically set by different mechanisms in many SUSY-breaking scenarios (see e.g. \cite{Giudice:1988yz, Witten:2001bf, Acharya:2006ia, Raby:2017ucc}). Therefore it is entirely possible to construct models which would give rise to such non-degeneracies.

\section{Gauge Coupling Unification}
\label{GCU.SEC}

So far we have mainly been discussing compatibility of high-scale supersymmetry with the Higgs boson mass measurement. We have found compatibility up to arbitrary high scales as long as the superpartners are allowed to have significant non-degeneracies among them. These very large non-degeneracies, however, are not ones that pull apart multiplets of GUT groups, such as the $\bar {\mathbf{5}}$ and $\mathbf{10}$ of $SU(5)$ or the $\mathbf{16}$ of $SO(10)$. Rather, the non-degeneracies are across these GUT representations, not within them. Strictly speaking this is not required, since orbifold GUTs, for example, can glue together several components of GUT representations into what looks like a single GUT representation from the IR point of view. Nevertheless, it is encouraging that even the simplest 
GUT theories pass the first test of compatibility with the Higgs boson mass and high-scale supersymmetry breaking.

A separate issue is whether exact gauge coupling unification indeed can happen in such theories. The lore is that low-scale supersymmetry is needed for exact gauge coupling unification. However, as we emphasized in \cite{Ellis:2015jwa}, exact gauge coupling unification is just as much an issue with the high-scale threshold corrections from GUT representation splittings as it is with low-scale threshold corrections that arise from superpartner thresholds. Thus, even the SM up to the high scale is compatible with gauge coupling unification from this perspective, although the corrections becomes quite large in that case, and one has to ask whether nature would rather have large corrections at the GUT scale for a SM GUT or very small corrections for a low-scale SUSY GUT. 

\begin{figure}[t]
\centering
\includegraphics[scale=0.45]{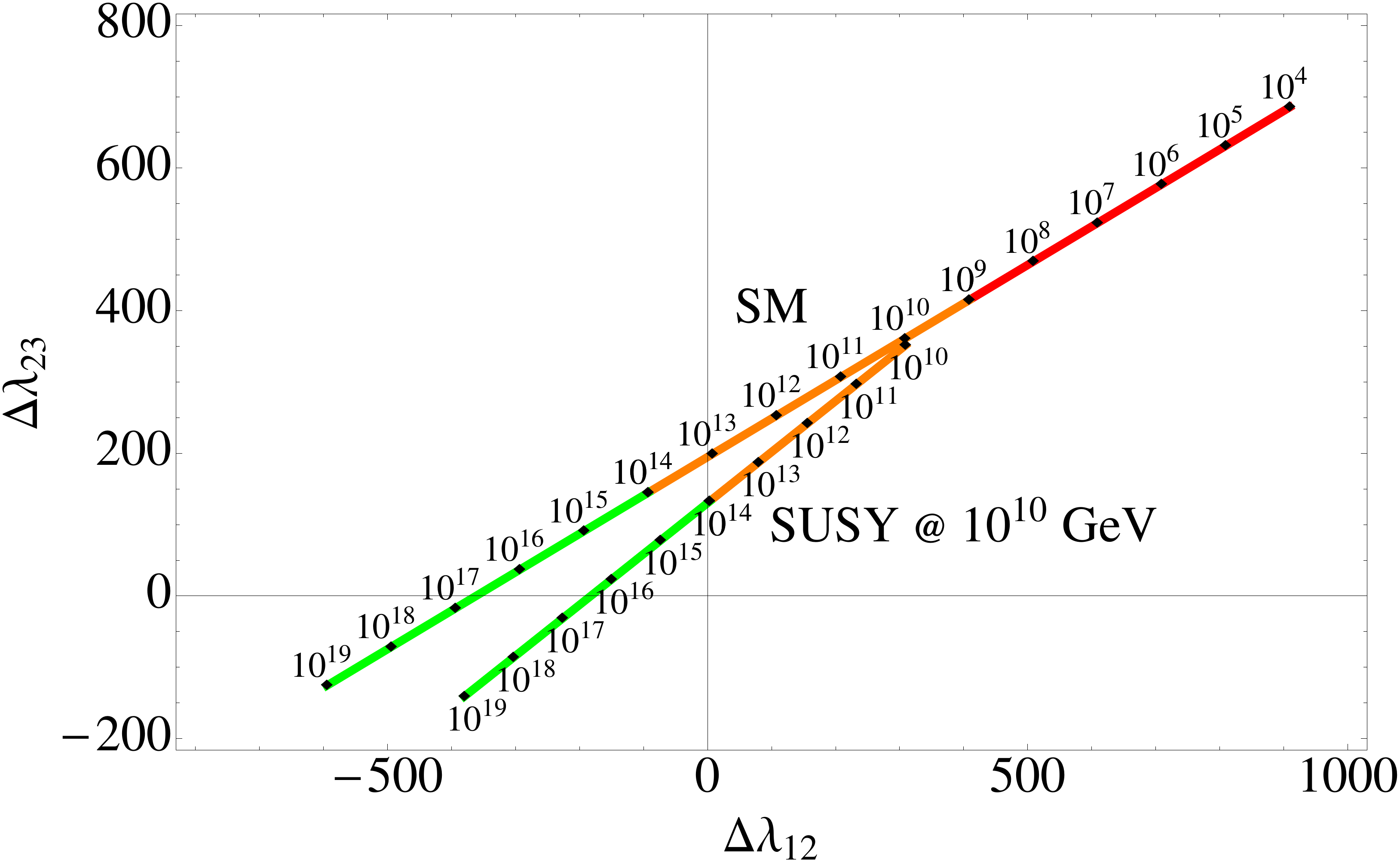}
\caption{Plot of the threshold corrections needed for exact gauge coupling unification. The numbers along the line are the scales $\mu_*$ at which the IR couplings are evaluated for unification and at which point the needed threshold corrections are computed and then plotted in the plane. The long straight line is assuming only the SM up to the highest scale. The second line that branches downward is for the case of superpartners existing at $10^{10}\gev$, which lowers the needed threshold corrections at high scales.}
\label{DeltaLambdaPlot_10to10.FIG}
\end{figure}

In the case of high-scale supersymmetry breaking, say the PeV scale or even the intermediate scale of $10^{10}\gev$, the situation is in between the SM GUT concerns and the very small corrections needed by low-scale supersymmetry to achieve exact gauge coupling unification. We can demonstrate this graphically using our visualization technique for the required threshold corrections to achieve exact gauge coupling unification. This is shown in Fig.~\ref{DeltaLambdaPlot_10to10.FIG}.

To understand the plot we need to build up the meaning of $\Delta\lambda_{23}$ and $\Delta\lambda_{12}$ (see \cite{Ellis:2015jwa} for more detailed discussion). First, we must pick a scale $\mu_*$ as a candidate GUT scale, at which point the required GUT threshold corrections for exact unification are computed. The values of $\mu_*$ are the numbers labeled along the lines in Fig.~\ref{DeltaLambdaPlot_10to10.FIG}. At $\mu_*$ we then compute the gauge couplings in the low-scale effective theory $g_i(\mu_*)$ and compare them to a candidate value of the GUT group's gauge coupling $g_U(\mu_*)$. These are related at one-loop~\cite{Weinberg:1980wa, Hall:1980kf} by
\beq
\label{BCunif.EQ}
\left(\frac{1}{g_i^2(\mu_*)}\right)_{\MS} = \left(\frac{1}{g_U^2(\mu_*)}\right)_{\MS} - \left(\frac{\lambda_i(\mu_*)}{48\pi^2}\right)_{\MS}
\eeq
where $\lambda_i(\mu_*)$ are the GUT-scale threshold corrections, specific to each gauge group coupling $g_i$ in the $\overline{MS}$ scheme.

From the low-energy point of view there are combinations of gauge couplings that do not involve the unification coupling $g_U(\mu_*)$
\begin{align}
\label{lij.EQ}
\left(\frac{\Delta\lambda_{ij}(\mu_*)}{48\pi^2}\right)_{\MS,~\DR} \equiv \left(\frac{1}{g_i^2(\mu_*)}-\frac{1}{g_j^2(\mu_*)}\right)_{\MS,~\DR} = \left(\frac{\lambda_j(\mu_*) - \lambda_i(\mu_*)}{48\pi^2}\right)_{\MS,~\DR}
\end{align}
for $i,j=1~,2,~3,~i\neq j$. These $\Delta\lambda_{ij}(\mu_*)$ are the horizontal and vertical axes in Fig.~\ref{DeltaLambdaPlot_10to10.FIG}.

Fig.~\ref{DeltaLambdaPlot_10to10.FIG} shows the required threshold corrections at various putative GUT scales $\mu_*$ for the SM and for intermediate scale supersymmetry, where the SUSY partners are all near $10^{10}\gev$. What we find is that supersymmetry deflects the ``thresholds line" corresponding to Eq. (\ref{lij.EQ}) to pass closer to the $(0,0)$ coordinates in the $(\Delta\lambda_{12},\Delta\lambda_{23})$ plane. It also increases the value of $\mu_*$ (i.e., GUT scale choice) that has its closest approach to $(0,0)$. The result is familiar: the introduction of supersymmetry both reduces the needed threshold corrections at the high-scale and increases the GUT scale (from the point of view of lowest-threshold correction is for higher values of $\mu_*$). This latter element is helpful since one generally requires that the GUT scale be above about $10^{15}\gev$ so that the $X,Y$ GUT gauge bosons do not induce too large dimension-six operators that cause the proton to decay faster than current limits allow. If supersymmetry existed at $\sim10^3\gev$, which is still compatible with constraints, the ``thresholds line" would pass very close to the $(0,0)$ point for $\mu_*\simeq 2\times 10^{16}\gev$, as is well-known and illustrated in Fig.~2 of \cite{Ellis:2015jwa}.

Since threshold corrections at the GUT scale can enable even pure SM theories to unify with exact gauge coupling unification~\cite{Lavoura:1993su, Ellis:2015jwa, Mambrini:2015vna}, the same must be true for $10^{10}\gev$ SUSY, since it only improves convergence to exact unification.  Compared to the SM threshold corrections $\Delta\lambda^{SM}$ for the GUT-scale choice $\mu_*=10^{16}\gev$, SUSY threshold corrections for a high-supersymmetry scale of $10^x\gev$ are given by $\sim \Delta\lambda^{SM}(x-3)/13$. More detailed analysis shows that this slightly overestimates the needed corrections for $6\lsim x\lsim 12$. Thus, exact gauge coupling unification is viable for intermediate values of supersymmetry breaking, which are also compatible with the Higgs boson mass constraint.

If superpartners were near the GUT scale, i.e. $\mS \sim 10^{15,16}$, gauge coupling unification would involve not only threshold corrections from the GUT boson/Higgs representations, but also from the superpartners themselves. Therefore, detailed analysis would be required to discuss precisely the conditions required for gauge coupling unification to occur, which we leave to future work.

\section{Conclusion}
\label{Conclusion.SEC}

The existence of supersymmetry at the weak scale has traditionally been assumed by reducing the naturalness problem of the quadratically sensitive Higgs sector of the SM. Our understanding of how naturalness is resolved in nature may be limited, and it is useful to consider theories of supersymmetry that are not beholden to the simplest notions of naturalness, and thus are not required to be at the weak scale. 

One important experimental prediction of minimal supersymmetry, even for superpartners at very high scales, is the existence of a relatively light Higgs boson. This has been seen by the LHC. We have shown in this paper that even arbitrary high scales of superpartners allow the light Higgs boson, due to the required matching of the SM effective theory Higgs self-interaction coupling to gauge couplings in the supersymmetric theory. Since gauge couplings stay perturbative up to the high scale this keeps the Higgs boson prediction light. Nevertheless, very high-scale supersymmetry above the PeV scale becomes increasingly difficult to reconcile with the Higgs mass; however, this may be achieved with large non-degeneracy factors, which we discussed in detail in the text. This has implications for the form of supersymmetry breaking that must be at play if supersymmetry is at very high scales well above the PeV scale. 

We re-iterate here that the ability to match the SM effective theory to a supersymmetric theory at high scales can be substantially altered by a change in the measurement of $m_t(m_t)$. Therefore an accurate measurement of this quantity would greatly improve the accuracy with which one could claim what conditions are necessary for matching to occur. For example, if $y_t(m_t)_{actual}=y_t(m_t)_{central}-3\sigma_{y_t(m_t)}$, the SM theory can be matched for totally degenerate spectra at all scales, whereas if $y_t(m_t)_{actual}=y_t(m_t)_{central}+3\sigma_{y_t(m_t)}$, the SM appears unlikely to be matched to GUT-scale SUSY (see Fig. \ref{16Vary.FIG}).

Another feature of supersymmetry, even for very high-scale values, is that requirements are reduced of large threshold corrections to achieve exact gauge coupling unification in a GUT. We have discussed how gauge coupling unification generally only improves in a supersymmetric theory, even at very high scales, with respect to the SM. Thus, we find that PeV scale~\cite{Wells:2003tf, ArkaniHamed:2004fb, Giudice:2004tc,  ArkaniHamed:2004yi, Wells:2004di} or intermediate scale supersymmetry~\cite{Hall:2013eko, Hall:2014vga}, two ideas that are prevalent in the literature for other reasons involving dark matter and neutrino physics, are compatible with the Higgs boson constraint and gauge coupling unification under the conditions described above.
\vspace{1cm}

\noindent
{\textit{\textbf{Acknowledgements:}}} We would like to thank E. Bagnaschi, J. Ellis, D. St\"ockinger, J. Quevillon and Z. Zhang for useful discussions. SARE and JDW are supported in part by DOE grant DE-SC0007859. SARE is also supported in part by a Rackham Research Grant. JDW is also supported in part by a Humboldt Fellowship. We thank the DESY theory group for their hospitality. 

\section*{Appendices}

\appendix

\section{Treatment of $\tan\beta$ in our analysis}
\label{TB.APP}

We discuss here the precise treatment of $\tan\beta$ in our analysis. Conventionally, $\tan\beta$ is defined as being
\begin{align}
 \tan\beta = \frac{\langle H_u \rangle}{\langle H_d \rangle}\ .
\end{align}
However, since we are considering SUSY scales significantly above the scale of electroweak symmetry breaking, the IR quantities $\langle H_u \rangle$ and $\langle H_d \rangle$ do not have meaning in the UV SUSY theory. Additionally, if we were to treat $\tan\beta$ according to the usual definition, this would complicate the analysis of the threshold corrections at the SUSY scale. Therefore, we define $\tan\beta$ as a scale-dependent tuneable parameter for the mixing angle from the two Higgs doublets in the UV theory, $H_u$ and $H_d$, to the two Higgs doublets in the IR theory, $H$ and $A$ (where $H$ is the light SM-like Higgs doublet).
\begin{align}
\begin{pmatrix} H \\ A \end{pmatrix} = \begin{pmatrix} \cos\beta & \sin\beta \\ -\sin\beta & \cos\beta \end{pmatrix} \begin{pmatrix} \tilde{H}_d \\ H_u \end{pmatrix} \ ,
\end{align}
where $\tilde{H}_d = -i \sigma_2 H_d$.
 As a result, $\tan\beta$ is defined as an input at the SUSY scale, and must be RGE evolved to different scales as in \cite{Sperling:2013eva}. This choice for the treatment of $\tan\beta$ is the same as that of Ref. \cite{Bagnaschi:2014rsa}.
 
\section{Corrections due to switching between the $\MS$ and $\DR$ schemes}
\label{Scheme.APP}

Here we summarize how switching between the $\MS$ and $\DR$ schemes affects the various couplings in the discussion in Section \ref{Setup.SEC}. 

The gauge couplings receive a finite correction:
\begin{align}
g_i^{\MS} = g_i^{\DR}\left\{ 1-\frac{g_i^2}{96\pi^2} C(G)\right\} \ ,
\end{align}
where $C(G)$ is the quadratic Casimir of the group $G$, defined as $C(G)\delta^{ab} \equiv f^{acd} f^{bcd}$, where $f^{abc}$ are the structure constants of the group. Thus for $U(1),~SU(N)$ gauge groups, $C(G) = 0,~N$ respectively.

The quartic coupling $\lambda_H$ also receives a finite correction due to the scheme switch:
\begin{align}
\lambda_H^{\MS} = \lambda_H^{\DR} - \frac{1}{16\pi^2}\left( \frac{9}{100}g_1^4 + \frac{3}{10}g_1^2 g_2^2 + \frac{3}{4}g_2^4 \right) \ ,
\end{align}
where the couplings on the right hand side are all in the $\DR$ scheme.
\section{One-loop SUSY matching to Higgs self-coupling}

The one-loop corrections to $\lambda^{tree}_{H}$ are as follows, with all couplings now shown in the $\MS$ scheme.
The one-loop corrections due to switching from $\DR$ to $\MS$ to match with the Standard Model are:
\begin{align}
\Delta\lambda_H^{(1),~scheme}(\mS) =  \frac{1}{16 \pi^2} \left\{-\frac{9}{100} g_1^4(\mS) - \frac{3}{10} g_1^2(\mS) g_2^2(\mS) - \left(\frac{3}{4} - \frac{\cos^2 2 \beta}{6}\right) g_2^4(\mS)\right\} \ .
\label{SchemeSwitch.EQ}
\end{align}
While most terms arise due to the correction of $\lambda_H^{\DR} \to \lambda_H^{\MS}$ as can be found, for example, in \cite{Bagnaschi:2014rsa}, the term proportional to $\cos^2 2 \beta$ arises due to the correction from $g_2^{\DR} \to g_2^{\MS}$ \cite{Bagnaschi:2014rsa}. More details on switching schemes can be found in Appendix \ref{Scheme.APP}.

The contribution from third generation squarks is given by
\begin{align}
\nonumber \Delta\lambda_H^{(1),~3rd~gen.}(\mS) &= \frac{1}{16\pi^2} \Bigg\{ 3y_t^2 \left( y_t^2 + \frac{1}{2} \left(g_2^2-\frac{1}{5}g_1^2 \right) \cos~ 2 \beta\right) \log\frac{\mStL^2}{\mS^2} 
\\ \nonumber &+ 3 y_t^2 \left( y_t^2 + \frac{2}{5} g_1^2 \cos~2\beta \right)\log\frac{\mStR^2}{\mS^2} 
\\ \nonumber &+\frac{\cos^2 2\beta}{300} \left( 3(g_1^4 + 25 g_2^4) \log \frac{\mStL^2}{\mS^2} + 24 g_1^4 \log \frac{\mStR^2}{\mS^2} + 6 g_1^4 \log\frac{\mSbR^2}{\mS^2}\right) 
\\ \nonumber&+ 6 y_t^4 \frac{(A_t-\mu \cot \beta)^2}{\mStL \mStR} \left( \tilde{F}_1\left(\frac{\mStL}{\mStR}\right)-\frac{1}{12}\frac{(A_t-\mu \cot \beta)^2}{\mStL \mStR} \tilde{F}_2\left(\frac{\mStL}{\mStR} \right)\right) 
\\ \nonumber &+ \frac{3}{4}y_t^2 \frac{(A_t-\mu \cot \beta)^2}{\mStL \mStR} \cos~2\beta \left( \frac{3}{5} g_1^2 \tilde{F}_3\left(\frac{\mStL}{\mStR} \right) + g_2^2 \tilde{F}_4\left(\frac{\mStL}{\mStR} \right)\right) 
\\ & - \frac{1}{4}y_t^2 \frac{(A_t-\mu \cot \beta)^2}{\mStL \mStR} \cos^2 2 \beta \left( \frac{3}{5} g_1^2 + g_2^2\right)\tilde{F}_5\left(\frac{\mStL}{\mStR} \right) \Bigg\}\ ,
\label{ThirdGenPart.EQ}
\end{align}
where the argument $(\mS)$ for all the couplings is implicit. The coupling $y_t$ is the top Yukawa coupling as defined in the SM phase, namely $y_t (m_t) = \sqrt{2}m_t(m_t) /v(m_t)$, where $m_t(m_t)$ is the pole top quark mass and $v(m_t)$ is the electroweak vev evaluated at the top mass threshold. We note that the last three lines are those that involve the stop $L-R$ mixing term $A_t - \mu \cot \beta$.

There are additional corrections from the sbottoms and staus not shown above, which are given in for example \cite{Lee:2015uza, Vega:2015fna}, which become important for very large values of $\tan\beta \gtrsim 50$. Since we restrict ourselves in our analysis to values $\tan\beta \leq 50$, we do not list these contributions here. Their inclusion does not alter the numerical results.

The contribution from the first two generations of squarks and sleptons is given by
\begin{align}
\nonumber \Delta\lambda_H^{(1),~1,2~gen.}(\mS) &= \frac{1}{16\pi^2} \Bigg\{ \frac{\cos^22\beta}{300}\sum_{i=1}^2 \Bigg( 3(g_1^4 + 25 g_2^4) \log \frac{\mSqL^2}{\mS^2} + 24 g_1^4 \log \frac{\mSuR^2}{\mS^2} \\ &+ 6 g_1^4 \log\frac{\mSdR^2}{\mS^2} + (9g_1^4 + 25 g_2^4) \log\frac{\mSlL^2}{\mS^2} + 18g_1^4 \log \frac{\mSeR^2}{\mS^2}\Bigg)  \Bigg\} \ ,
\label{OtherGenPart.EQ}
\end{align}
for the parts explicitly dependent on the squark and slepton masses. The scalar corrections to $\lambda_H$ also yield a mass-independent correction which depends only on $g_i$ and $\tan\beta$, which we denote as $\Delta\lambda_H^{(1),~s^24\beta}(\mS)$:
\beq
\Delta\lambda_H^{(1),~s^24\beta}(\mS) = -\frac{1}{16\pi^2}\frac{3}{16}\left( \frac{3}{5}g_1^2 + g_2^2\right)\sin^2 4\beta \ .
\label{SinBeta.EQ}
\eeq
The corrections due to the heavy Higgs boson doublet $A$ is
\begin{align}
\nonumber \Delta\lambda_H^{(1),~A}(\mS) &= \frac{1}{16\pi^2} \Bigg\{ \frac{1}{4800}\Bigg( 261 g_1^4 + 630 g_1^2 g_2^2 + 1325 g_2^4 - 4 \cos~4\beta (9g_1^4 + 90 g_1^2 g_2^2 + 175 g_2^4) \\&- 9 \cos~8\beta(3g_1^2 + 5g_2^2)^2 \Bigg) \log\frac{m_A^2}{\mS^2}\Bigg\} \ ,
\label{mAPart.EQ}
\end{align}
while the correction due to Gauginos/Higgsinos is given by
\begin{align}
\nonumber \Delta\lambda_H^{(1),~-ino}(\mS) &= \frac{1}{16\pi^2} \Bigg\{  \frac{1}{400} \Bigg(-36g_1^4 G_1(x_1) - 120 g_1^2 g_2^2 G_2(x_1, x_2) - 100g_2^4 G_3(x_2) \\ \nonumber &+ 80 \lambda_H( g_1^2 ~G(x_1) + 5 g_2^2 ~G(x_2))  \\ 
\nonumber&+ \cos4\beta \Big( -36g_1^4 G_4(x_1) + 120 g_1^2 g_2^2 G_5(x_1, x_2) -100 g_2^4 G_6(x_2)\Big) \\
\nonumber&+ 16 \sin 2 \beta \Big( g_1^2(6g_1^2 - 5 \lambda_H) \tilde{F}_5(x_1) +5 g_2^2\left[4 g_1^2~G_7(x_1,x_2) + (10 g_2^2 - 5 \lambda_H) \tilde{F}_5(x_2)\right]\Big)
\Bigg) \\ &- \frac{1}{6}\cos^2 2\beta \left( 2 g_2^4 \log \frac{M_2^2}{\mS^2} + \left(\frac{9}{25}g_1^4 + g_2^4\right)\log\frac{\mu^2}{\mS^2}\right) \Bigg\} \ 
\label{Gauginos1L.EQ}
\end{align}
where $x_a = M_a / \mu$.

The loop functions appearing in the expressions above are defined to be
\begin{align}
G_1(x) &= \frac{-1-8x^2+7x^4+2x^6+x^2(-3-11x^2+2x^4)\log x^2}{(x^2-1)^3}\ , \\ 
G_2(x,y) &= \frac{x^3(1+2x^2)\log x^2}{(x-y)(x^2-1)^2} - \frac{(y^2-1)2x^2 - 2y^2 - 3xy -1}{(x^2-1)(y^2-1)} - \frac{y^3(1+2y^2)\log y^2}{(x-y)(y^2-1)^2} \ , \\
G_3(x) &= \frac{-3-26x^2 +25x^4 + 4x^6 + x^2(-9-35x^2+8x^4)\log x^2}{(x^2-1)^3} \ , \\
G_4(x) &= \frac{1+2x^2-3x^4+x^2(3+x^2) \log x^2}{(x^2-1)^3} \ , \\
G_5(x,y) &= \frac{x^3 \log x^2}{(x-y)(x^2-1)^2} + \frac{1+xy}{(x^2-1)(y^2-1)} - \frac{y^3 \log y^2}{(x-y)(y^2-1)^2} \ , \\
G_6(x) &= \frac{3+4x^2-5x^4 - 2x^6+x^2(9+x^2+2x^4)\log x^2}{(x^2-1)^3} \ , \\
G_7(x,y) &= \frac{3}{2}\left(\frac{x+y}{(x^2-1)(y^2-1)} + \frac{x^4 \log x^2}{(x^2-1)^2(x-y)}-  \frac{y^4 \log y^2}{(y^2-1)^2(x-y)} \right)\ ,
\end{align}
\begin{align}
G(x) &= \frac{-3(x^4-6x^2+1)}{2(x^2-1)^2}+\frac{3x^4(x^2-3)\log x^2}{(x^2-1)^3}\, \\
\tilde{F}_1(x) &= \frac{x \log x^2}{x^2-1}\, \\
\tilde{F}_2(x) &= \frac{6x^2(2-2x^2+(1+x^2)\log x^2)}{(x^2-1)^3}\, \\
\tilde{F}_3(x) &= \frac{2x(5(1-x^2) + (1+4x^2) \log x^2}{3(x^2-1)^2}\, \\
\tilde{F}_4(x) &= \frac{2x(x^2-1-\log x^2)}{(x^2-1)^2}\, \\
\tilde{F}_5(x) &= \frac{3x(1-x^4+2x^2 \log x^2)}{(1-x^2)^3}\, 
\end{align}

\bibliographystyle{apsrev}
\bibliography{dissertationBIB}

\end{document}